\documentclass[a4paper,11pt]{article}
\usepackage{jheppub}
\usepackage{graphicx}
\usepackage{amsmath, amssymb} 
\usepackage{subcaption}
\usepackage[dvipsnames]{xcolor}
\title{\boldmath Holographic metals at finite volume}
\author{Lucas Acito,} 
\author{Nicol\'as Grandi,} 
\affiliation{{\it Instituto de F\'\i sica La Plata (IFLP), CONICET  \& \\
Departmento de F\'\i sica  Dr. Emil H. Bose, UNLP\\
C.C. 67, (1900) La Plata, Argentina.}}
\emailAdd{acitolucas@gmail.com} 
\emailAdd{grandi@fisica.unlp.edu.ar} 
   
\abstract{
We construct the electron star solution in asymptotically global AdS spacetime, and investigate its thermodynamic stability properties, both globally with respect to the Reissner-Nordström and thermal AdS spacetimes, and under local perturbations of the thermodynamic variables. {\color{black} The solution is sourced with a perfect fluid at finite temperature, which implies that the ``suspended electron cloud'' previously reported in the literature does not appear as an equilibrium phase.} We interpret the resulting system as that of a holographic metal confined to a finite volume. {\color{black} The corresponding {\color{black}grand canonical phase diagram} reveals a rich structure of transitions involving three dominant phases, which depend strongly on the fluid parameters characterizing the star. Finally, we identify a quantum critical point at finite chemical potential, around which these different phases are organized.}
}
\begin{document}
\maketitle
\section{Introduction}
One of the most interesting applications of holography over the past decade has been the description of strongly coupled condensed matter systems. In this context, considerable effort has been devoted to exploring the properties of superconducting \cite{Hartnoll:2008,Faulkner:2010,Gubser:2010} and metallic \cite{Hartnoll:2004,Faulkner:2011,Cubrovic:2009,Hartnoll:2011,Hartnoll:2011b} phases from a holographic perspective.

A central feature of many holographic descriptions of condensed matter phenomena is that the boundary field theory has no intrinsic scale. Consequently, when constructing the {\color{black}grand canonical phase diagram}, one thermodynamic quantity (\emph{e.g.}, the chemical potential $\mu$) must set the overall energy scale. This restricts the remaining quantities (such as the temperature $T$ and the magnetic field $B$) to form the independent dimensionless parameters of the {\color{black}grand canonical phase diagram} (\emph{i.e.}, $T/\mu$ and $B/\mu$). This scaling behavior contrasts with that of real high-$T_c$ materials, which have an intrinsic dimensionful parameter $a$ that dictates the physical scale, allowing for independent thermodynamic ratios such as $T/a$, $B/a$, and $\mu/a$.
To our knowledge, two main approaches have been proposed in the literature to overcome the aforementioned limitation. The first involves introducing an additional bulk field whose boundary value defines the scale $a$ \cite{Kiritsis:2015hoa}. However, this faces us to the new problem of finding a sensible physical interpretation for the new degree of freedom from the boundary point of view. The second proposal is to abandon the planar setup predominantly used in holographic literature, in favor of a spherical boundary \cite{Verlinde:2011, Arsiwalla:2011}. In this scenario, the boundary curvature radius naturally plays the role of the scale $a$.  

{\color{black}
The metallic phase in the holographic theory is described by adding fermionic matter to the bulk in the form of a perfect fluid coupled to the gravitational field. The solution corresponding to a neutral fluid with a spherical boundary is called a \emph{holographic neutron star} \cite{Verlinde:2011, Arsiwalla:2011}. On the other hand, the configuration sourced by charged matter and having a planar boundary is dubbed \emph{electron star} \cite{Hartnoll:2011,Hartnoll:2011b}.}
The aim of this work is to investigate the {\color{black}grand canonical phase diagram}, parametrized by the dimensionless temperature $T/a$ and chemical potential $\mu/a$, of an electron star with a spherical boundary. Holographically, this setup corresponds to a strongly coupled metallic phase confined within a spherical vessel. 

{\color{black} Holographic neutron stars at finite temperature have been studied in \cite{Grandi:2018, Grandi:2020} (see also \cite{Canavesi:2023}), where the gravitating perfect fluid was described by a Fermi-Dirac distribution function \cite{Arguelles:2013, Arguelles:2014, Ruffini:2015}. For electron stars, finite-temperature effects were introduced in \cite{Hartnoll:2010, Thorlacius:2011} by means of a black hole horizon, while the perfect fluid itself was kept at zero temperature. In this setup, a cloud of charged particles appears above the horizon, balanced by the competing electromagnetic and gravitational forces. However, such a cloud is expected to disappear when the fluid is brought into equilibrium with the black hole. Indeed, a fluid at finite temperature contains a fraction of its particles moving fast enough to escape cloud and reach the horizon. These particles are then swallowed by the black hole, generating a continuous particle flow that would eventually dissipate the cloud. In this work, we checked that this is indeed the case, and the particle cloud is not present among the relevant phases of an electron star at finite temperature.}
\section{The model}
\label{sec:dynamics}
We consider the dynamics of a charged perfect fluid coupled to the gravitational and electromagnetic fields in $3+1$ dimensions (for details, see Appendix A). The relevant degrees of freedom are the metric $g_{\mu\nu}$, the electromagnetic field $A_\mu$, and the fluid four-velocity $u^\mu$.
The equations of motion take the form
\begin{align}
\label{eq:EOM}
R_{\mu\nu}-\frac{1}{2}R\,g_{\mu\nu}-\frac{3}{L^2}g_{\mu\nu}=\kappa^2\left(T^{\sf EM}_{\mu\nu}+T^{\sf Fluid}_{\mu\nu}\right)\,,
&  \qquad\qquad&  \nabla_\mu F^{\nu\mu}=e^2 J^\nu \, ,
\end{align}
where $T^{\sf EM}_{\mu\nu}$ and $T^{\sf Fluid}_{\mu\nu}$ are the electromagnetic and fluid energy-momentum tensors, respectively, and $J^\mu$ is the electric current
\begin{align}
    &T^{\sf EM}_{\mu\nu}= \frac{1}{e^2}\left(F_{\mu\alpha}F_{\,\,\nu}^{\alpha}-\frac{1}{4}g_{\mu\nu}F_{\alpha\beta}F^{\alpha\beta}\right)\ , &J^\mu=\sigma u^\mu\ , \nonumber\\
    &T^{\sf Fluid}_{\mu\nu}= g_{\mu\nu}P+(\rho+P) u_{\mu} u_{\nu}\,.
\end{align}
These expressions are given in terms of the pressure $P$, the energy density $\rho$, and the electric charge density $\sigma$. Explicit forms for these quantities can be obtained in the $mL \gg 1$ limit, which corresponds to a large number of particles within one AdS radius, by using the Thomas-Fermi approximation. In other words, we use the statistics of a large ensemble of charged fermions in local thermodynamic equilibrium.

In the context of finite-temperature field theory, the system is described by a grand canonical ensemble where particles and antiparticles constitute distinct thermal excitations within the Fock space~\cite{Kapusta:2006}. Both species contribute additively to the energy density and pressure, whereas the charge density is determined by the net particle number ({\em i.e.}, the difference between particle and antiparticle populations)~\cite{Glendenning:1997}. Consequently, the equations of state take the form
\begin{equation}
    \begin{split}
        \rho&=\frac{g}{8\pi^3}\int \left(f(p,\mu)+f(p,\bar\mu)\right)\sqrt{p^2+m^2}\:d^3p\ ,
        \\
        P&=\frac{g}{8\pi^3}\int \left(f(p,\mu)+f(p,\bar\mu)\right)\frac{p^2}{\sqrt{p^2+m^2}}\:d^3p\ ,
        \\
        \sigma&=\frac{g}{8\pi^3}\int \left(f(p,\mu)-f(p,\bar\mu)\right)\:d^3p \ ,
        \\
    \end{split}
\end{equation}
where $f(p,\mu)$ is the distribution function, and $\mu$ and ${\bar \mu}$ denote the particle and antiparticle chemical potentials, respectively. The quantity $g$ counts the number of fermionic species. Here, we have assumed a particle charge of $+1$ (following the minimal coupling convention detailed in Appendix A).

Focusing on static and stable electron star configurations, we impose the equilibrium condition $\bar{\mu} = -\mu$. This condition implies a locally balanced creation and annihilation of particle-antiparticle pairs but does not imply symmetry in the number densities ($n \neq \bar{n}$)~\cite{Landau:1980}. Instead, the system sustains a net conserved charge density $\sigma$ determined by the value of $\mu$. Substituting $\bar{\mu} = -\mu$ and transforming to the dimensionless energy variable $\epsilon = E/m$, the equations of state yield
\begin{equation}\label{eq:thermodynamics}
    \begin{split}
        \tilde{\rho}&\equiv L^2\kappa^2\rho=\gamma\int_1^\infty \left(f(\epsilon,\tilde{\mu})+f(\epsilon,-\tilde{\mu})\right)\,\epsilon^2\sqrt{\epsilon^2-1}\:d\epsilon\ ,
        \\
        \tilde{P}&\equiv L^2\kappa^2P=\frac{\gamma}{3}\int_1^\infty \left(f(\epsilon,\tilde{\mu})+f(\epsilon,-\tilde{\mu})\right)\,(\epsilon^2-1)^{3/2}\:d\epsilon\ ,
        \\        
        \tilde{\sigma}&\equiv L^2\kappa e\,\sigma=\frac{\gamma}{\tilde{m}}\int_1^\infty \left(f(\epsilon,\tilde{\mu})-f(\epsilon,-\tilde{\mu})\right)\,\epsilon\sqrt{\epsilon^2-1}\:d\epsilon\ ,
    \end{split}
\end{equation}
where we have introduced the rescaled dimensionless quantities $\tilde\rho$, $\tilde P$ and $\tilde\sigma$, and the dimensionless parameters $\gamma = g L^2 \kappa^2 m^4/(2\pi^2)$ and $\tilde{m} = m\,\kappa/e$. The function $f(\epsilon, \tilde{\mu})$ is the Fermi-Dirac distribution
\begin{equation}
    f(\epsilon,\tilde{\mu})=\frac{1}{e^{\frac{\epsilon-\tilde{\mu}}{\tilde{T}}}+1}\ ,
\end{equation}  
where we have defined the dimensionless temperature $\tilde{T}=T/m$ and chemical potential $\tilde{\mu}=\mu/m$. These quantities are spatial functions satisfying the conditions of thermodynamic equilibrium to be discussed below.

Note that we are working not only in the limit $mL\gg 1$ but also in the classical limit $\kappa^2/L^2\ll1$, both of which enter the definition of $\gamma$. However, as discussed in \cite{Hartnoll:2011}, we are interested in a scaling limit where $\gamma\sim1$. This is achieved by taking $e^{-1}\sim \kappa/L\ll1$; i.e., we require the gravitational coupling to be proportional to the square of the Maxwell coupling (the `probe brane' limit). Furthermore, these limits imply that the scaled mass is of order unity, $\tilde{m}\sim1$.

\bigskip

We use a static and spherically symmetric ansatz for the background fields, given by
\begin{align}
\label{eq:Ansatz}
&A=\frac{eL}{\kappa}h\:dt\ ,
&u=u^0\partial_t\ ,
\nonumber\\
&ds^2=L^2\left(-f\:dt^2+g\:dr^2+r^2d\Omega_2^2\right)\ ,&
\end{align}
where $d\Omega_2^2$ is the metric of the two-sphere, and $f$, $g$, and $h$ are functions of the radial coordinate $r$. The multiplicative constants in the gauge field ansatz have been chosen for later convenience. The fluid four-velocity must satisfy the timelike condition $g_{\mu\nu}u^\nu u^\mu=-1$, which yields $u^0=1/(L\sqrt{f})$.
In what follows, we find it convenient to parameterize the metric functions as
\begin{align}
    f&=e^{\chi}\left(1-\frac{2M}{r}+\frac{Q^2}{2r^2}+r^2 \right)\ ,
    \\
    g&=\left(1-\frac{2M}{r}+\frac{Q^2}{2r^2}+r^2 \right)^{-1}\ ,
\end{align}
in terms of the new radial functions $M$, $\chi$, and $Q$, where $Q$ represents the conserved charge associated with the $U(1)$ symmetry written in a locally Minkowskian frame~\cite{Bekenstein:1971}
\begin{equation}
    Q=e^{-\frac{\chi}{2}}\:r^2\:h'\,.
\end{equation}
We also identify the electric field in a locally Minkowskian frame as
\begin{equation}
    E=\sqrt{g^{tt}}\sqrt{g^{rr}}F_{rt}=e^{-\frac{\chi}{2}}h'
    \,.
\end{equation}
Notice that $E=Q/r^2$. In terms of these new variables, the equations of motion for the system read
\begin{equation}\label{eq:eom_star}
    \begin{split}
    \chi'&= r g (\tilde{P}+\tilde{\rho})\ ,
    \\
    E'&=-\frac{2E}{r} +\sqrt{g}\:\tilde{\sigma}\ ,
    \\
    M'&= \frac{r^2}2\left(\tilde{\rho}+r\:E\sqrt{g}\:\tilde{\sigma}\right)\ .
    \end{split}
\end{equation}
From the conservation of the energy-momentum tensor we can obtain another equation
\begin{equation}   
    \tilde{P}'= \frac{2E^2}{r} +E\,E' + \frac{\chi'\left(g'-g\:\chi'\right)}{2\:r\:g^2} \,,
\end{equation}
This equation is redundant, as the pressure  $\tilde{P}(r)$ is fully determined by the equations of state \eqref{eq:thermodynamics}. Nevertheless, it serves as a valuable consistency check.

Necessary and sufficient conditions for our charged perfect fluid to be in thermodynamic equilibrium are given by the Tolman and Klein relations when the particles are coupled to an external field (for details, see Appendix A and refs.~\cite{Landau:1980,Shi:2021})
\begin{equation}\label{eq:thermo_equilibrium}
    \tilde{T}(r)=\frac{\tilde{T}_0}{\sqrt{f(r)}} \qquad\text{and}\qquad 
    \tilde{\mu}(r)=\frac{h(r)}{\tilde{m}\sqrt{f(r)}}\,,
\end{equation}
where $\tilde{T}_0$ is a reference value.
Note that when the electromagnetic field is constant we recover the holographic neutron star conditions of \cite{Grandi:2018,Grandi:2020}.

The equations of motion \eqref{eq:eom_star} must be solved subject to the equations of state \eqref{eq:thermodynamics} under the thermodynamic equilibrium conditions \eqref{eq:thermo_equilibrium}. In the absence of a horizon, initial conditions must be set at $r=0$. To obtain a regular solution, we require three initial conditions for the three first order equations
\begin{equation}
    M(0)=\chi(0)=E(0)=0 \,. 
    \label{eq:boundary.conditions}
\end{equation}
Notice that this implies $Q(0)=0$ and then $f(0)=g(0)=1$. Defining further $h(0)=\tilde m\:\tilde \mu_0$, these conditions identify the reference value $\tilde{T}_0$ in \eqref{eq:thermo_equilibrium} and $\tilde{\mu}_0$ as the central temperature and the central chemical potential, respectively. Once the fluid parameters $\gamma$ and $\tilde m$ are fixed, the set of solutions is indexed by the central values $\tilde{T}_0$ and $\tilde{\mu}_0$.

On the other hand, when a horizon $r_0$ is present, initial conditions should be set at the horizon, as $f(r_0)=0=1/g(r_0)$. We also require the gauge potential $h$ to vanish at the horizon $h(r_0)=0$ to avoid a conical singularity in the Euclidean continuation~\cite{Kobayashi:2007}.

%
\section{Solutions: black hole, thermal AdS, and electron star}\label{sec:solutions}

In the presence of a horizon, the solution of the equations of motion \eqref{eq:eom_star} is consistent with $\tilde{\sigma}=\tilde{\rho}=\tilde{P}=0$. This corresponds to the absence of a fluid throughout the entire radial domain. By solving the remaining equations, we obtain
\begin{equation}\label{eq:bh_sol}     
    \chi=0\,,\quad M=M_0\,, \quad Q=Q_0 \quad\text{and}\quad h=-\frac{Q_0}{r}+\mu_{\text{\tiny BH}}\,. 
\end{equation}
This yields the Reissner-Nordström-AdS black hole (RNBH) solution, where $M_0$ and $Q_0$ denote the black hole mass and charge, respectively, with a singularity at the origin $r=0$. The condition  $h(r_0)=0$ thus implies $\mu_{\text{\tiny BH}}=Q_0/r_0$, where $r_0$ is the outer horizon radius.

Outside the horizon, one might think of a possible configuration consisting of a cloud of charged particles surrounding the black hole. This type of star configurations have been studied in \cite{Hartnoll:2010,Thorlacius:2011}. There, the temperature of the particle cloud was fixed to zero, implying that the cloud is not in thermodynamic equilibrium with the black hole.  If instead we consider the cloud at a non-zero temperature, we find that these configurations are not stable. This happens because the exponential tail in the fluid density profile reaches the horizon, generating a continuous flow of particles being swallowed by the black hole. In conclusion, for a thermal star with a black hole around the origin, the only thermodynamically stable configuration arises when all particles eventually fall into the black hole, resulting in a pure Reissner-Nordström   solution. This is explained in more detail Appendix C. 

\bigskip

Alternatively, the equations of motion admit a regular vacuum solution known as `Thermal AdS$_4$' (TAdS). This geometry is obtained by setting $M_0=Q_0=0$ in Eq. \eqref{eq:bh_sol} and fixing the gauge potential to a constant value, $h(r)=\mu_{\text{\tiny TAdS}}$, throughout the bulk. This solution represents a regular vacuum state with a constant chemical potential extending from the boundary to the origin. { This is consistent with boundary conditions \eqref{eq:boundary.conditions}.}

\bigskip

{ To obtain the electron star solution, we solve the equations of motion \eqref{eq:eom_star} alongside the equations of state \eqref{eq:thermodynamics} with boundary conditions \eqref{eq:boundary.conditions} for specific central values $(\mu_0,T_0)$. Since the equations of motion \eqref{eq:eom_star} imply $E'(0)=M'(0)=\chi'(0)=0$, a naive numerical integration, as for example the Euler method, would result in a trivial solutions where all the variables remain constant for all values of $r$. A successful numerical integration then requires that we move away from $r=0$ to a small non-vanishing radius $r_\epsilon$ by using a Frobenius expansion. The electron star solutions for larger values of $r$ are then obtained numerically. In our calculations, we used a \texttt{Wolfram Mathematica} code~\cite{Github}.} 

\begin{figure}
	\begin{center}
	\includegraphics[width=0.33\textwidth]{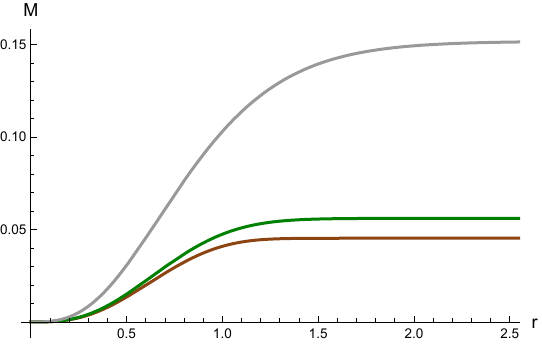}%
    \includegraphics[width=0.33\textwidth]{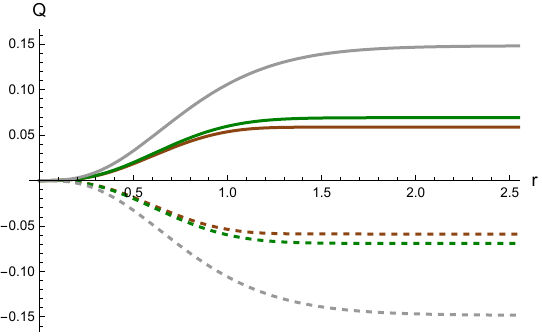}%
    \includegraphics[width=0.33\textwidth]{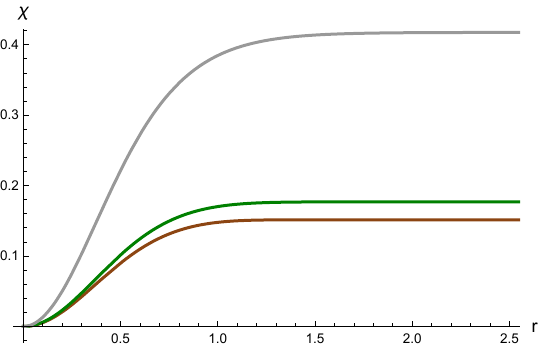}
		\caption{Radial profiles of the local mass $M(r)$, charge $Q(r)$, and metric function $\chi(r)$ for the Electron Star, obtained via the shooting method by varying the central parameters $\tilde{\mu}_0$ and $\tilde{T}_0$. The colors of the curves correspond to the specific points marked in the phase space diagram (Fig.~\ref{fig:criticality}). Solid lines represent solutions with positive central chemical potential $\tilde{\mu}_0$, while dashed lines correspond to $-\tilde{\mu}_0$ (with the same magnitude). Note that the mass and $\chi$ profiles coincide for both signs, illustrating the invariance of the geometry under charge conjugation.}
        \label{fig:MQchi}
	\end{center}
\end{figure}

By varying the central values $\tilde{T}_0$ and $\tilde{\mu}_0$, we integrated the equations of motion \eqref{eq:eom_star} and obtained the mass and charge profiles shown in Fig.~\ref{fig:MQchi}. We observe that the star exhibits a well-defined boundary within the numerical radial cutoff, beyond which the radial functions saturate to constant values
\begin{equation}
    M(r)\sim M_s\,, \qquad Q(r)\sim Q_s \,,\qquad\chi(r)\sim\chi_s\,,
\end{equation}
impying that the solution takes the asymptotic form
\begin{equation}\label{star_asymp}
    f(r)\sim e^{\chi_s}\left(1-\frac{2M_s}{r}+\frac{Q_s^2}{2r^2}+r^2\right) \qquad\text{and}\qquad h(r)\sim -e^{\frac{\chi_s}2}\frac{Q_s}{r}+\tilde m\:\left(\tilde\mu_0 +\Delta \tilde\mu\right) \,.
\end{equation}
where $\Delta\tilde\mu$ is an additional constant obtained from the numerical integration. Consequently, the metric at large $r$ approaches the Reissner-Nordström-AdS black hole solution, with its time coordinate rescaled by a factor of $e^{\frac{\chi_s}{2}}$.

We are interested in the asymptotic thermodynamic values in order to compare our results with other solutions characterized by the same temperature and chemical potential. Thus, we require the solution to approach pure $AdS_4$ at the boundary. Since the lapse function $f(r)$ in \eqref{star_asymp} asymptotically approaches the $AdS_4$ form, eq.~\eqref{eq:thermo_equilibrium} yields 
\begin{equation}\label{eq:T_mu_asymtp}
    \tilde{\mu}\sim \frac{\tilde{\mu}_\infty}{\sqrt{1+r^2}} \qquad\text{and}\qquad \tilde{T}\sim \frac{\tilde{T}_\infty}{\sqrt{1+r^2}}\,,
\end{equation}
where we defined
\begin{equation}\label{eq:muT_map}
    \tilde{T}_\infty=e^{-\frac{\chi_s}{2}}\,\tilde{T}_0 \qquad\text{and}\qquad 
    \tilde{\mu}_\infty=e^{-\frac{\chi_s}{2}}\, \left(\tilde\mu_0 +\Delta \tilde\mu\right) \,.
\end{equation}
{Then the asymptotic parameters $(\tilde{\mu}_\infty,\tilde{T}_\infty)$  are related to the central ones $(\tilde\mu_0,\tilde T_0)$ by means of the numerical method. A critical choice affecting the precision of this relation is that of the order of the aforementioned Frobenius expansion. We employ a 10th-order expansion; by comparing the extracted parameters with those obtained from an 11th-order expansion, we find that the relative errors are of $\mathcal{O}\left(10^{-12}\right)$, ensuring that the numerical uncertainties are well under control. Regarding the numerical integration from $r_\epsilon$ on, the introduced error is much smaller, of ${\cal O}(10^{-25})$.}

\begin{figure}
    \vspace{.8cm}
    \centering
    \begin{minipage}[c]{0.6\textwidth}
        \centering
        \includegraphics[width=\linewidth]{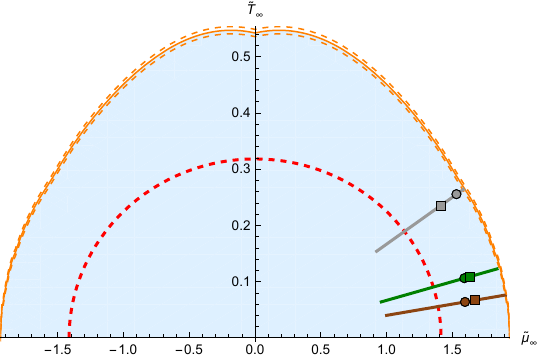}
    \end{minipage}
    \hfill
    \begin{minipage}[c]{0.39\textwidth}
        \vspace{0.05cm} 
        \includegraphics[width=\linewidth]{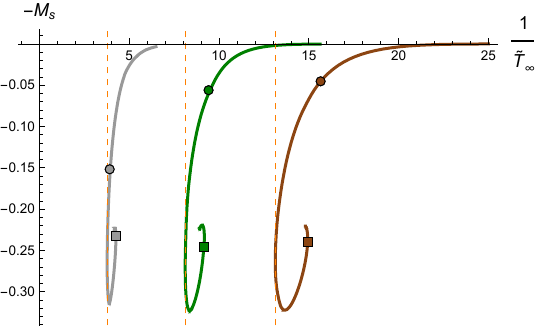}
    \end{minipage}
    \\~\\~\\
    \centering
    \includegraphics[width=0.33\linewidth]{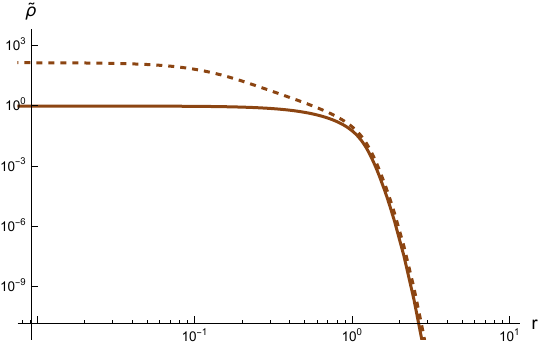}%
    \hfill
    \includegraphics[width=0.33\linewidth]{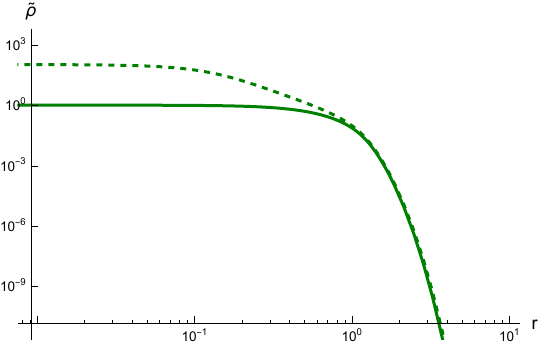}%
    \hfill
    \includegraphics[width=0.33\linewidth]{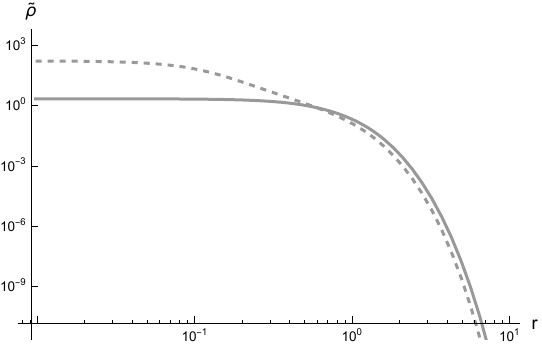}
    \vspace{.7cm}
    \caption{\underline{Top left}: The shaded region indicates the domain where the electron star solution exists. This region is delimited by the orange critical curve constructed by identifying the critical point on the corresponding Katz curve. {\color{black} The dashed orange lines indicate the uncertainty bounds of the critical curve, derived from the dispersion of the extracted critical points.} The three colored lines represent families of solutions with fixed $\tilde{\mu}_\infty/\tilde{T}_\infty$, where circles and squares denote representative {\color{black}thermodynamically} stable and unstable configurations, respectively. For reference, the Hawking-Page curve is shown as a dashed red line (see Sec.~\ref{sec:phase_transitions}). \underline{Top right}: Katz {\color{black}thermodynamic} stability curves for the three families indicated in the {\color{black}grand canonical phase diagram}. The vertical dashed orange lines mark the critical point obtained for each curve, while the circles and squares denote the stable and unstable configurations shown in the bottom panel. \underline{Bottom}: Log-log plot for the radial density profiles corresponding to the specific configurations marked in the two top panels; the colors match the phase space trajectories. Solid lines represent {\color{black}thermodynamically} stable solutions (circles in the top figures), characterized by an abrupt edge profile, while dashed lines correspond to unstable solutions (squares) exhibiting power-law behavior.}
    \label{fig:criticality}
\end{figure}

We construct the star profiles in terms of $\tilde{T}_\infty$ and $\tilde{\mu}_\infty$ by shooting from the central values $\tilde{T}_0$ and $\tilde{\mu}_0$ to the boundary, as shown in Fig.~\ref{fig:criticality}. The critical curve is constructed using the Katz criterion~\cite{Katz:1978,Katz:2002}, which  identifies a vertical asymptote at which the slope changes sign (or, in other words, a turning point) in the caloric curve $-M_s$ {\em vs.} $1/\tilde{T}_\infty$ while keeping $\tilde{\mu}_\infty/\tilde{T}_\infty$ fixed. Thermodynamical stability can only change at such turning points, where the system develops a zero mode. As the Katz curve spirals inward, each subsequent turning point introduces an additional negative mode. Consequently, the first {\color{black}thermodynamically} unstable branch possesses exactly one 
unstable mode, and the system never regains stability once it enters the spiraling region.  

In the holographic neutron star case~\cite{Canavesi:2023}, the ratio $\tilde \mu/\tilde T$ is constant throughout the bulk, meaning that fixing the central parameters $\tilde{\mu}_0/\tilde{T}_0$ is equivalent to fixing the boundary values. This facilitates the calculation of the Katz instability curves, as plotting the curve $-M_s$ as a function of $1/\tilde T_\infty$ at constant $\tilde\mu_\infty/\tilde T_\infty$ is equivalent to plotting it at constant $\tilde \mu_0/\tilde T_0$. In the present electron star case, however, the ratio varies radially, such that $\tilde{\mu}_0/\tilde{T}_0\neq \tilde{\mu}_\infty/\tilde{T}_\infty$. 
Then, the values of $-M_s$ calculated at constant  $\tilde \mu_0/\tilde T_0$ sit above a curve in the $(1/\tilde{T}_\infty,\tilde{\mu}_\infty/\tilde{T}_\infty)$ plane, which is determined by the map  \eqref{eq:muT_map}. Varying the value of $\tilde \mu_0/\tilde T_0$ moves this curve, and the corresponding values of $-M_s$ span a surface above the $(1/\tilde{T}_\infty,\tilde{\mu}_\infty/\tilde{T}_\infty)$ plane.
%
%
From the corresponding interpolation we can extract contours of constant $\tilde{\mu}_\infty/\tilde{T}_\infty$, yielding the Katz curves shown in Fig.~\ref{fig:criticality}. 
%
%
The critical boundary is then formed by the set of critical points, each identified by a vertical asymptote in the corresponding curve $-M_s$ as a function of $1/\tilde T_\infty$.
{\color{black} The interpolation procedure required to construct the Katz curves constitutes the main source of uncertainty in determining this critical boundary. To account for this, we define the final critical curve by averaging the extracted critical points and computing the absolute error around the mean curve, yielding a value of $\mathcal{O}(10^{-3})$ as illustrated in the figure.} 

As seen in Fig.~\ref{fig:criticality}, according to the Katz criterion, {\color{black}thermodynamically} stable and unstable solutions coexist in the $\tilde{\mu}_\infty$ {\em vs.} $\tilde{T}_\infty$ plane only within the domain delimited by the critical (orange) curve, although they correspond to different central values. Beyond that region, no electron star solutions exist. In terms of the central parameters, stable solutions correspond to low values of $\tilde{\mu}_0$ and $\tilde{T}_0$, while the unstable configurations emerge as these parameters are increased. Furthermore, the {\color{black}thermodynamically} stable density profiles shown in the figure correspond exclusively to `pure core' stars, which are characterized by an abrupt edge. Conversely, the unstable solutions exhibit a power-law (no-scale) behavior at the edge of the profile, eventually evolving into a `cusp' star. 

\begin{figure}
    \vspace{1.7cm}
    \centering
    \begin{subfigure}[b]{0.48\textwidth}
        \centering
        \includegraphics[width=\textwidth]{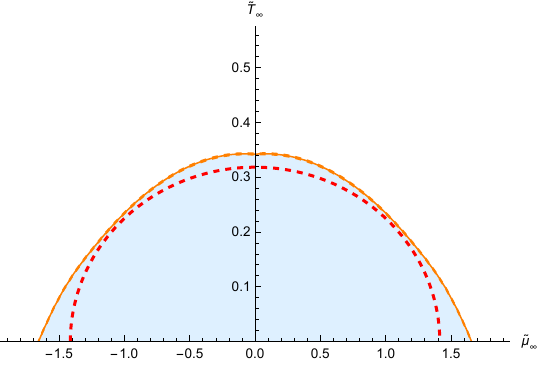}
        \caption{$\gamma = 11$ and $\tilde{m}=1$}
        \label{fig:critA}
    \end{subfigure}
    \hfill
    \begin{subfigure}[b]{0.48\textwidth}
        \centering
        \includegraphics[width=\textwidth]{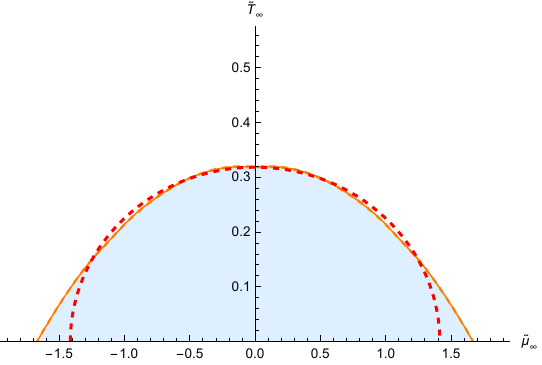}
        \caption{$\gamma = 16$ and $\tilde{m}=1$}
        \label{fig:critB}
    \end{subfigure}

    \vspace{1cm}

    \begin{subfigure}[b]{0.48\textwidth}
        \centering
        \includegraphics[width=\textwidth]{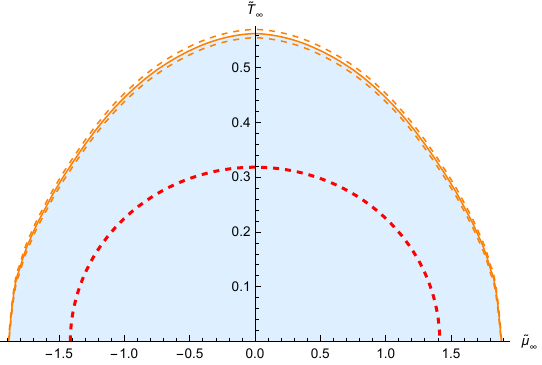}
        \caption{$\gamma = 1$ and $\tilde{m}=1.07$}
        \label{fig:critC}
    \end{subfigure}
    \hfill
    \begin{subfigure}[b]{0.48\textwidth}
        \centering
        \includegraphics[width=\textwidth]{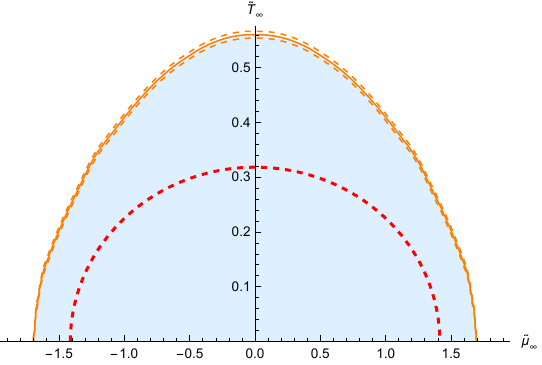}
        \caption{$\gamma = 1$ and $\tilde{m}=2$}
        \label{fig:critD}
    \end{subfigure}
    \vspace{1cm}
    \caption{Critical curves (solid orange lines) {\color{black}with their associated uncertainty bounds (dashed orange lines)} and the domain of existence for the electron star solutions (shaded regions), determined according to the Katz criterion, corresponding to different values of $\gamma$ and $\tilde m$. All plots share the same scale for direct comparison. For reference, the Hawking-Page transition is indicated by the dashed red line (see Sec.~\ref{sec:phase_transitions}).}
    \label{fig:criticality2}
\end{figure}

Upon varying the fluid parameters $\tilde m$ and $\gamma$, the possible configurations remain restricted to pure core (stable) and cusp (unstable) stars. However, as shown in Fig.~\ref{fig:criticality2}, the critical curve itself shrinks vertically as $\gamma$ increases (Figs.~\ref{fig:critA} and \ref{fig:critB}) and horizontally as $\tilde{m}$ increases (Figs.~\ref{fig:critC} and \ref{fig:critD}). Above the red dotted line, the energetically dominant solution is the Reissner-Nordstr\"om black hole, while below it is the thermal AdS background  (see Sec.~\ref{sec:phase_transitions}). Notably, in Fig.~\ref{fig:critB}, the electron star domain crosses this line, indicating that beyond the electron star phase, we could find a TAdS phase instead of a black hole one. 

\bigskip 
Having established the domain of existence for the electron star, we must now determine its thermodynamic stability relative to the competing vacuum solutions, namely the Reissner-Nordström-AdS black hole and Thermal AdS. To this end, in the following section, we evaluate and compare the free energies of these geometries to construct the complete {\color{black}grand canonical phase diagram} of the system.




\section{Grand canonical phase diagram}\label{sec:phase_transitions}
Let us now compute the thermodynamic potential in the grand canonical ensemble for the different geometries obtained in the previous section. The grand canonical potential is obtained from the partition function via the relation
\begin{equation}\label{eq:F_def}
    \Omega=-T\log Z \,.
\end{equation}
The problem then reduces to computing the partition function from the gravitational path integral in Euclidean signature,
\begin{equation}
    Z=\int Dg \:D\Phi\:\: e^{-S_E[g,\Phi]} \,,
\end{equation}
with
\begin{equation}
    S_E[g,\Phi]=\frac{1}{2\,\kappa^2} \int \sqrt{g}\left(R[g]+\dots\right)+S_{\sf b.t.} \,,
\end{equation}
where $\Phi$ denotes all the matter fields and $S_{\sf b.t.}$ represents the boundary terms. We work in the saddle-point approximation, where the path integral is dominated by the classical gravity solutions, {\em i.e.}, the on-shell configurations (we denote the Euclidean on-shell action by $I$). In this semiclassical approximation, we must sum over all relevant saddle points contributing to the path integral
\begin{equation}
    Z_{\sf Grav}\approx e^{-I^{\sf(Star)}} +e^{-I^{\sf(BH)}} +e^{-I^{\sf(TAdS)}}\,.
\end{equation}
The details of each Euclidean on-shell action computation are provided in Appendix B. Note that, since the action is of order $1/\kappa^2$ (which is large in the classical limit), the partition function is exponentially dominated by the term with the smallest action ({\em i.e.}, the minimum grand canonical potential). This implies that we get $\Omega=TI$  for each saddle. We defined the temperature and chemical potential in the previous section ensuring that all three saddle points share the same boundary values $\tilde{T}_\infty$ and $\tilde{\mu}_\infty$. Thus, by expressing the on-shell actions in terms of $\tilde{T}_\infty$ and $\tilde{\mu}_\infty$, we can map the phase plane to determine the dominant phase. The resulting potentials to be compared are given by
\begin{equation}\label{eq:free_energies}
    \begin{aligned}
        &\tilde{\Omega}^{\sf(BH)}= \frac{\pi}{27}\left(2\pi\,\tilde{T}_\infty-\sqrt{2} \sqrt{3\,\tilde{\mu}_\infty^2 +8\pi^2\,\tilde{T}_\infty^2-6}\right) \left(\sqrt{2} \sqrt{3\,\tilde{\mu}_\infty^2+8 \pi^2\,\tilde{T}_\infty^2-6}+4 \pi\, \tilde{T}_\infty\right)^2
        \\
        &\tilde{\Omega}^{\sf(TAdS)}= 0
        \\
        &\tilde{\Omega}^{\sf(Star)}= 4\pi\,e^{\frac{\chi_s}{2}}\left(2 M_s -Q_s\,\tilde{\mu}_\infty-e^{-\chi_s}\int_0^\infty dr\,\,r^2 e^{\frac{\chi}{2}}\left(\tilde{\rho}+\tilde{P}-\tilde{m}\,\tilde{\mu}\,\tilde{\sigma}\right)\right) \,, 
    \end{aligned}
\end{equation}
where we have defined $\tilde{\Omega}=({\kappa^2}/{L^2})\Omega$. Note that these grand canonical potentials are symmetric under $\tilde{\mu}_\infty\leftrightarrow-\tilde{\mu}_\infty$ (as expected for a relativistic fluid of fermions, where the equation of state \eqref{eq:thermodynamics} shares this symmetry); we will thus focus on the $\tilde{\mu}_\infty\geq0$ sector of the {\color{black}grand canonical phase diagram}.
{\color{black} 
To estimate the numerical uncertainty of the star potential $\tilde{\Omega}^{\sf (Star)}$, we evaluate the error associated with each of the terms in \eqref{eq:free_energies}, by comparing the values computed using Frobenius expansions of the 10th and 11th order (see section \ref{sec:solutions}).  This results in a final relative error of $\mathcal{O}(10^{-9})$ for the star potential.}


\paragraph{Quantum critical point.} It is instructive to first treat the zero-temperature case separately and analyze the quantum critical point. The equations of motion \eqref{eq:eom_star} remain unchanged, but the Fermi-Dirac distribution becomes a Heaviside step function centered at $\vert{}\tilde{\mu}\vert{}=1$  \cite{Verlinde:2011}. Focusing on the $\tilde{\mu}>0$ case (where only particles are present), the equations of state become
\begin{equation}
        \tilde{\rho}= \gamma\int_1^{\tilde{\mu}} \epsilon^2\sqrt{\epsilon^2-1}\:d\epsilon,\qquad
        \tilde{\sigma}= \frac{\gamma}{\tilde{m}}\int_1^{\tilde{\mu}}  \epsilon\sqrt{\epsilon^2-1}\:d\epsilon \qquad\&\qquad \tilde{P}=\tilde{m}\,\tilde{\mu}\,\tilde{\sigma}-\tilde{\rho}\,,      
\end{equation} 
with $\tilde{\mu}\geq1$. Using the numerical integration to map the central chemical potential $\tilde{\mu}_0$ to the boundary value $\tilde{\mu}_\infty$, we compute the free energies in \eqref{eq:free_energies}.

\begin{figure}
    \centering
    \includegraphics[width=0.49\linewidth]{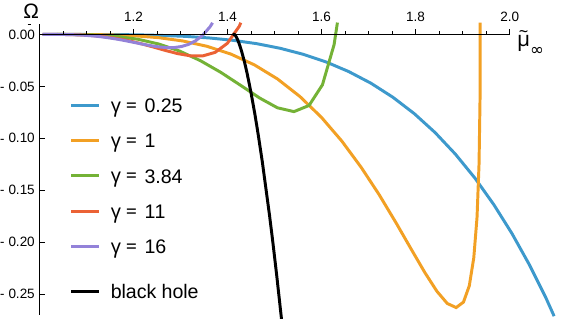}%
    \hfill\includegraphics[width=0.49\linewidth]{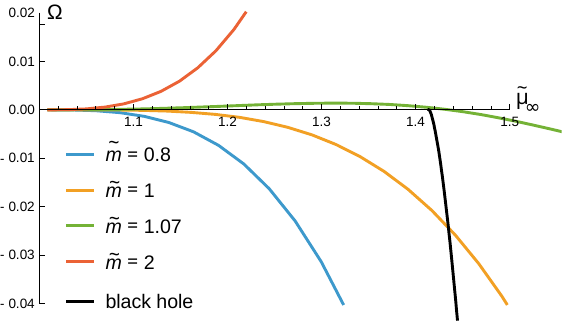}
    \caption{Grand canonical potential plots at $\tilde{T}_{\infty} = 0$ for different choices of the fluid parameters, compared against the black hole and TAdS phases (the latter lying on the horizontal axis). \underline{Left}: The fluid parameter $\tilde{m}$ is fixed at $\tilde{m} = 1$ while $\gamma$ is varied. A triple point is obtained at $\gamma \approx 11$, whereas for $\gamma > 11$, an intermediate TAdS phase emerges between the electron star and the black hole. For $\gamma \approx 3.84$, the quantum critical point between the electron star and the black hole reaches a maximum. \underline{Right}: The parameter $\gamma$ is fixed at $\gamma = 1$ while $\tilde{m}$ is varied. A triple point is observed at $\tilde{m} \approx 1.07$; for $\tilde{m} < 1.07$, an intermediate electron star phase arises between the TAdS and black hole phases. In this case, the maximum $\tilde{\mu}_\infty$ for the quantum critical point between the electron star and the black hole corresponds to the limit $\tilde{m} \to 0$.}
    \label{fig:FT0}
\end{figure}

\begin{figure}
    \centering
    \includegraphics[width=0.6\linewidth]{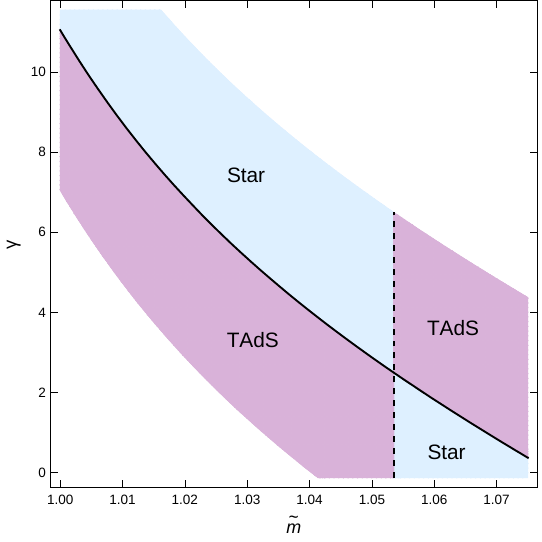}
    \caption{Fluid parameter space $(\tilde{m}, \gamma)$ evaluated at $\tilde{T}_{\infty} = 0$. The solid black line represents the locus of triple points at $\tilde{T}_{\infty} = 0$ and $\tilde{\mu}_\infty = \sqrt{2}$. Moving away from it yields different dominant solutions, depending on the chosen values for the fluid parameters (see Fig.~\ref{fig:FT0}). The white region represents unexplored values of the fluid parameters.}
    \label{fig:mgamma}
\end{figure}

In Fig.~\ref{fig:FT0}, we show the resulting grand canonical potentials for the holographic star across various values of the fluid parameters, comparing them against the free energies of the TAdS (horizontal axis) and the black hole (black curve). 
A star profile can only exist if the central potential satisfies $\tilde{\mu}_0 \geq 1$. This in turn implies that the boundary chemical potential $\tilde\mu_\infty$ has a lower bound, below which there is no star and the TAdS solution dominates. As $\tilde\mu_\infty$ grows, two scenarios are possible depending on the values of the fluid parameters $\tilde m$ and $\gamma$. For large enough $\tilde m$ the star never dominates, and the TAdS solution is replaced by the black hole when the chemical potential reaches the Hawking-Page value, which in our conventions is $\sqrt2$ (see the right panel of Fig.~\ref{fig:FT0}). If $\tilde m$ is instead small and $\gamma$ sits below some critical value, we may have a star with negative energy which dominates over the TAdS as the chemical potential grows, and is later replaced by the black hole. Alternatively for larger $\gamma$ the system re-enters the TAdS phase after leaving the star and before reaching the black hole (see the left panel of Fig. \ref{fig:FT0}). This behavior is summarized in the {\color{black}grand canonical phase diagram} of Fig.~\ref{fig:mgamma}, where the colored regions indicate whether the electron star or TAdS is dominant at $\tilde{\mu}_\infty=\sqrt{2}$. These regions are interchanged at $\tilde{m}\approx1.053$. 

Interestingly, our numerical results indicate the existence of a triple point where the electron star grand canonical potential becomes negative precisely at Hawking-Page transition $\tilde\mu_\infty=\sqrt2$.
As seen in the left panel of Fig.~\ref{fig:FT0}, for $\tilde{m}=1$ and $\gamma\approx11$ the electron star curve crosses the horizontal axes exactly at the point where the black hole appears. Analogously, in right panel the triple point appears at $\gamma\approx 1$ and $\tilde m\approx 1.07$. This triple point is depicted by the black line in Fig.~\ref{fig:mgamma}. 

{\color{black} In this calculation, numerical uncertainties in $\tilde \Omega^{\sf Star}$ propagate to the phase boundary, resulting in a relative error in the critical value of $\tilde{\mu}_\infty$ of order ${\cal O}(10^{-9})$.}

\newpage 

\begin{figure}
    \vspace{.5cm}
    \centering
    \begin{minipage}[c]{0.641\textwidth}
        \centering
        \includegraphics[width=\linewidth]{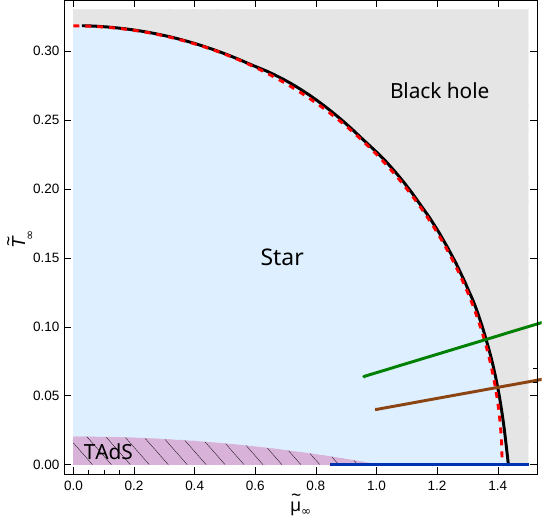}
    \end{minipage}
    \hfill
    \begin{minipage}[c]{0.35\textwidth}
        \centering
        \includegraphics[width=\linewidth]{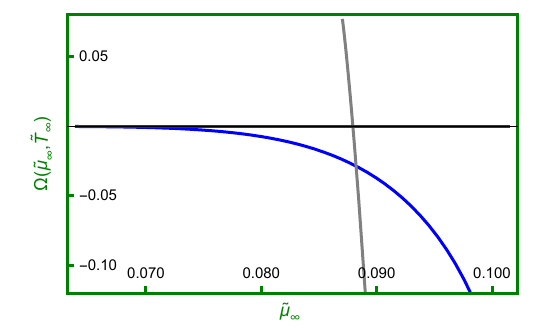}
        \vspace{0.05cm} 
        \includegraphics[width=\linewidth]{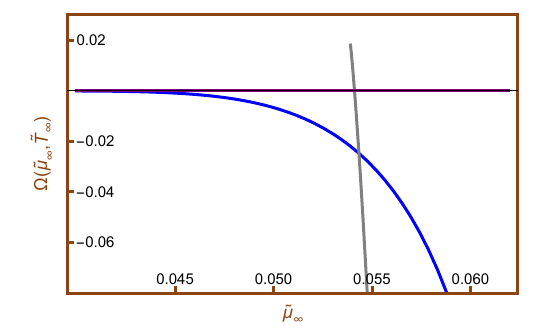}
        \vspace{0.05cm} 
        \includegraphics[width=\linewidth]{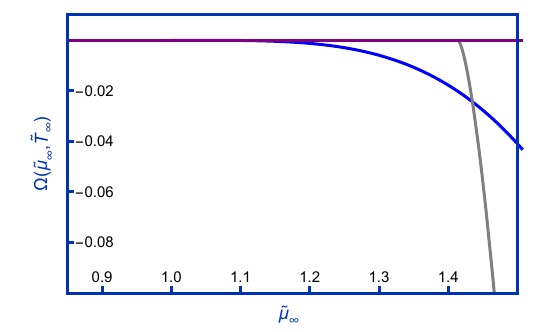}
    \end{minipage}
    \vspace{.5cm}
    \caption{\textbf{Left}: {\color{black}Grand canonical phase diagram} of the holographic electron star in the boundary plane $(\tilde{\mu}_\infty,\tilde{T}_\infty)$. The light blue region indicates where the electron star is the dominant solution (minimal grand canonical potential), while the purple and grey region correspond to TAdS and black hole dominance, respectively. The TAdS region is hatched because numerical limitations prevent its direct observation at finite temperatures. Nevertheless, at $\tilde{T}_{\infty} = 0$ we do observe the dominance of TAdS over the electron star. Since this region can be seen with other fluid parameters, we expect it to be present here as well. The solid black lines denote first-order phase transition, and the dashed red line indicates where the Hawking-Page transition would occur in the absence of the star~\cite{Chamblin:1999}. The colored lines (green, brown, blue) represent the slices along which we evaluate the free energies shown in the right panels. \textbf{Right}: grand canonical potential comparisons for the electron star (blue), black hole (grey), and TAdS (purple) corresponding to the cuts in the {\color{black}grand canonical phase diagram}. The frame colors match the cuts in the left panel. The black dots mark the phase transition points. The blue frame corresponds to the $\tilde{T}_\infty=0$ case, from which the quantum critical point is determined. }
    \label{fig:phase.diagram}
\end{figure}

\paragraph{{\color{black}Grand canonical phase diagrams}.} We now consider the finite-temperature electron star profiles obtained in the previous section and compare their grand canonical potential at each point in the $(\tilde{\mu}_\infty,\tilde{T}_\infty)$ plane with those of the competing saddles. This construction is shown in Fig.~\ref{fig:phase.diagram}, where we identify the regions of dominance corresponding to the minimal grand canonical potential. It is important to remark that we evaluated the free energies within the thermodynamic domain bounded by the star's critical curve (see Fig.~\ref{fig:criticality}). Consequently, while the star solution technically exists in regions overlapping with the TAdS and black hole phases, the star phase is dominant only within a bounded region preceding its collapse.

\begin{figure}
    \vspace{1cm}
    \centering
    \begin{subfigure}[b]{0.48\textwidth}
        \centering
        \includegraphics[width=\textwidth]{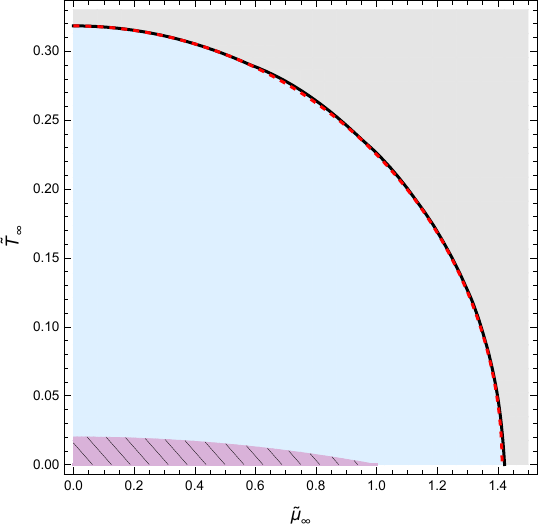}
        \caption{$\gamma = 0.25$}
        \label{fig:gamma0p25}
    \end{subfigure}
    \hfill
    \begin{subfigure}[b]{0.48\textwidth}
        \centering
        \includegraphics[width=\textwidth]{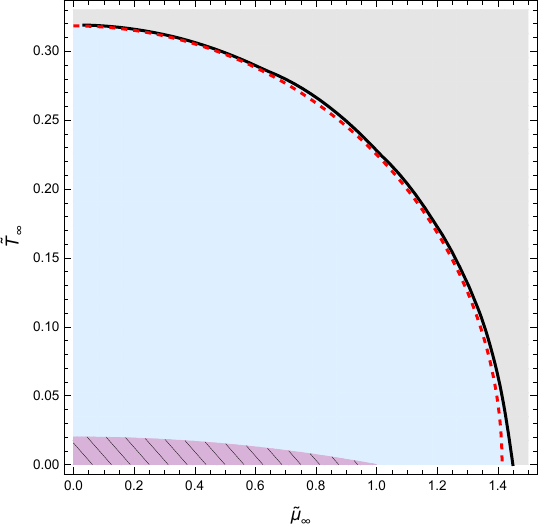}
        \caption{$\gamma = 3.84$}
        \label{fig:gamma3p84}
    \end{subfigure}

    \vspace{0.5cm}

    \begin{subfigure}[b]{0.48\textwidth}
        \centering
        \includegraphics[width=\textwidth]{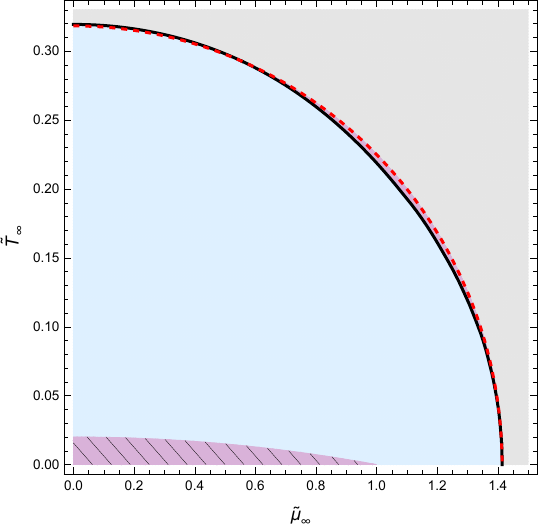}
        \caption{$\gamma = 11$}
        \label{fig:gamma11}
    \end{subfigure}
    \hfill
    \begin{subfigure}[b]{0.48\textwidth}
        \centering
        \includegraphics[width=\textwidth]{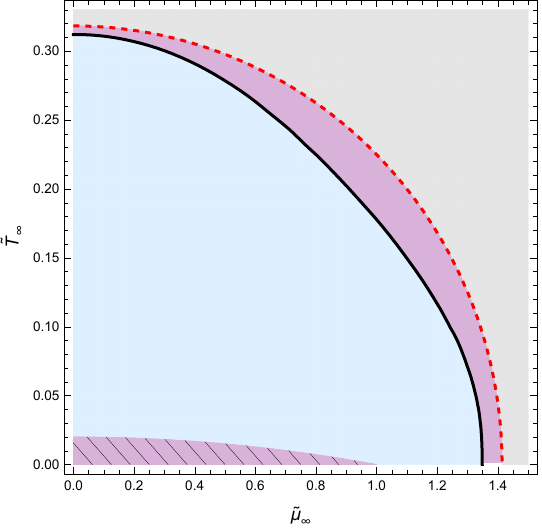}
        \caption{$\gamma = 16$}
        \label{fig:gamma16}
    \end{subfigure}
    \caption{{\color{black}Grand canonical phase diagrams} of the holographic electron star keeping fixed one of the fluid parameter, $\tilde{m} = 1$, and varying the other parameter $\gamma$ as indicated in the respective panels. The diagrams are plotted in the boundary plane $(\tilde{\mu}_\infty, \tilde{T}_\infty)$. The light blue region indicates where the electron star is the dominant solution (minimal grand canonical potential), whereas the purple and grey regions correspond to the TAdS and black hole phases, respectively. The hatched regions remain unexplored at finite temperatures due to numerical limitations; however, their existence is inferred from $\tilde{T}_{\infty} = 0$ observations and results obtained for other values of the fluid parameters. Solid black lines denote first-order phase transitions involving the electron star, while the dashed red line indicates the Hawking-Page transition.}
    \label{fig:pd.gamma.fix}
\end{figure}

From the {\color{black}grand canonical phase diagram} in Fig.~\ref{fig:phase.diagram}, we observe sharp boundaries where the thermodynamic dominance switches between solutions, indicating first-order phase transitions. For the fluid parameters $\tilde{m}=\gamma=1$ at finite temperatures, there are two dominant phases—the electron star and the black hole—separated by a phase transition at $\Omega^{\sf(Star)}=\Omega^{\sf(BH)}$. Based on the zero-temperature analysis, a TAdS region is expected to emerge at low temperatures for $\tilde{\mu}_\infty \lesssim  1$. However, we were unable to numerically resolve this region at finite $\tilde{T}_{\infty}$ due to precision limitations. Physically, near the transition between the electron star and the TAdS phase, charge repulsion overcomes gravitational attraction, causing the core of the star to become very diluted, orders of magnitude smaller than machine precision. However, because the zero-temperature analysis confirms the dominance of the TAdS phase for small enough $\tilde{\mu}_\infty$, and because the TAdS region is explicitly observed for other parameter choices that enlarge its domain (see Fig.~\ref{fig:pd.gamma.fix}), we infer its existence here and denote it with a hatched region. Finally, the star / black hole boundary can be compared to the standard Hawking-Page phase transition between TAdS and a charged black hole~\cite{Chamblin:1999}, represented by the dashed line at $\Omega^{\sf(TAdS)}=\Omega^{\sf(BH)}$.

\begin{figure}
    \vspace{1cm}
    \centering
    \begin{subfigure}[b]{0.48\textwidth}
        \centering
        \includegraphics[width=\textwidth]{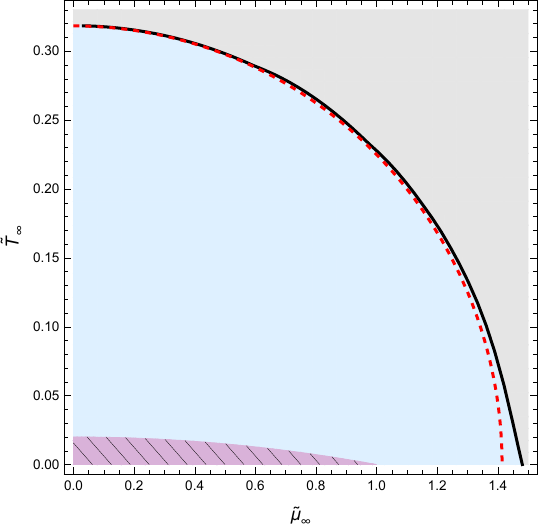}
        \caption{$\tilde{m} = 0.8$}
        \label{fig:m0p8}
    \end{subfigure}
    \hfill
    \begin{subfigure}[b]{0.48\textwidth}
        \centering
        \includegraphics[width=\textwidth]{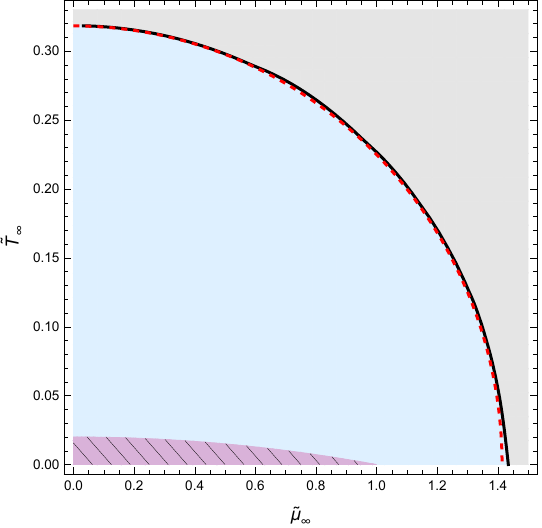}
        \caption{$\tilde{m} = 1$}
        \label{fig:m1}
    \end{subfigure}

    \vspace{0.5cm}

    \begin{subfigure}[b]{0.48\textwidth}
        \centering
        \includegraphics[width=\textwidth]{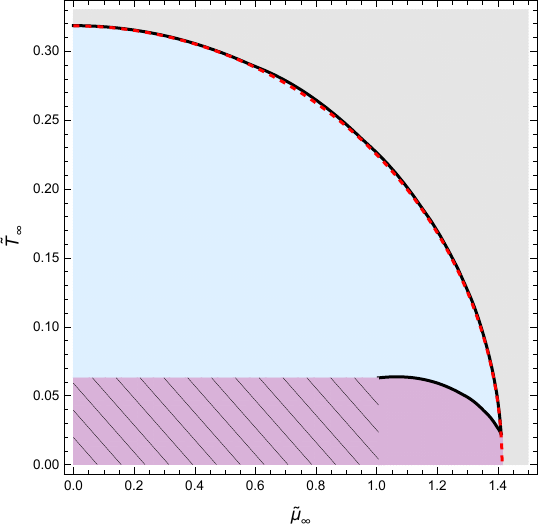}
        \caption{$\tilde{m} = 1.07$}
        \label{fig:m1p07}
    \end{subfigure}
    \hfill
    \begin{subfigure}[b]{0.48\textwidth}
        \centering
        \includegraphics[width=\textwidth]{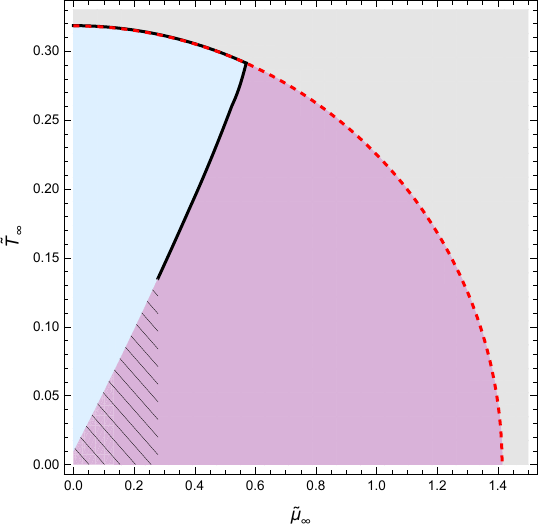}
        \caption{$\tilde{m} = 2$}
        \label{fig:m2}
    \end{subfigure}
    \caption{{\color{black}Grand canonical phase diagrams} of the holographic electron star keeping fixed one of the fluid parameter, $\gamma = 1$, and varying the other parameter $\tilde{m}$ as indicated in the respective panels. The diagrams are plotted in the boundary plane $(\tilde{\mu}_\infty, \tilde{T}_\infty)$. The light blue region indicates where the electron star is the dominant solution (minimal grand canonical potential), whereas the purple and grey regions correspond to the TAdS and black hole phases, respectively. The hatched regions remain unexplored at finite temperatures due to numerical limitations; however, their existence is inferred from $\tilde{T}_{\infty} = 0$ observations and results obtained for other values of the fluid parameters. Solid black lines denote first-order phase transitions involving the electron star, while the dashed red line indicates the Hawking-Page transition.}
    \label{fig:pd.m.fix}
\end{figure}

As we vary the fluid parameters, the {\color{black}grand canonical phase diagram} becomes richer, as shown in Figs.~\ref{fig:pd.gamma.fix} and \ref{fig:pd.m.fix}. In these plots, we used the same values as in the zero-temperature analysis of Fig.~\ref{fig:FT0}. Note that in Figs.~\ref{fig:gamma0p25}, \ref{fig:gamma3p84}, \ref{fig:m0p8}, and \ref{fig:m1}, the electron star phase collapses into a black hole; this transition consistently occurs beyond the Hawking-Page curve. The quantum critical points coincide with our $\tilde{T}_\infty=0$ analysis. While the critical values can increase depending on the fluid parameters, the transition curves asymptotically approach the Hawking-Page curve at higher temperatures and smaller chemical potentials. If $\gamma$ is increased while keeping $\tilde{m}$ fixed, a TAdS phase emerges between the electron star and the black hole. Depending on the exact value, this can result in a triple point among the three phases (Fig.~\ref{fig:gamma11}), or three entirely separate regions featuring an electron star-TAdS transition followed by a standard Hawking-Page transition (Fig.~\ref{fig:gamma16}). Interestingly, in Fig.~\ref{fig:gamma11} a second triple point arises for higher temperatures at $\tilde T_\infty\approx 0.5$ and $\tilde\mu_\infty\approx0.3$. On the other hand, setting $\gamma=1$ and increasing $\tilde{m}$ enlarges the internal TAdS region (Figs.~\ref{fig:m1p07} and \ref{fig:m2}), yielding a triple point in both cases. Finally, the transition curves involving the electron star (solid black lines) are drawn up to the limit where numerical precision allows for the reliable computation of the star profiles. Overall, these finite-temperature {\color{black}grand canonical phase diagrams} are in complete agreement with the features established at zero temperature. 

\newpage 

{\color{black}

The present results agree with the published literature in the corresponding limits. In particular,  the neutral limit studied in \cite {Verlinde:2011, Arsiwalla:2011,Grandi:2018, Grandi:2020, Canavesi:2023} is easily recovered as $e\to0$, as the charge density $\sigma$ decouples. In our variables, this corresponds to $\tilde m\to \infty$, indicating that the neutral star only dominates at small chemical potential, see Fig.~\ref{fig:m2}. Moreover, in this region the star is stable, which matches the results of \cite{Grandi:2018, Grandi:2020, Canavesi:2023}. The charged spherical star of \cite{Arsiwalla:2011} is recovered in the $\tilde T_\infty\to0$ limit of our setup. In Fig.~\ref{fig:pd.m.fix} we see that, as the fermion charge $e$ is increased at fixed fermion mass (\emph{i.e.} as the variable $\tilde{m}$ decreases) the critical chemical potential at zero temperature grows, coinciding with the results reported in \cite{Arsiwalla:2011}.
The charged planar zero temperature limit of \cite{Hartnoll:2011} is recovered by a double scaling limit, in which the charge and the mass of the star go to infinity \cite{Arsiwalla:2011}.  Finally, as explained above (see section \ref{sec:solutions} and Appendix \ref{app:electron_cloud}) the ``suspended electron cloud" phase of Refs. \cite{Hartnoll:2010, Thorlacius:2011}  is not recovered as a limit of our construction, since it is a non-equilibrium state. 

As in the zero temperature case, the numerical uncertainties propagate to the {\color{black}grand canonical phase diagrams} resulting in relative errors on both the critical temperature $\tilde{T}_\infty$ and chemical potential $\tilde{\mu}_\infty$ of order $\lesssim {\cal O}(10^{-9})$.
}

\section{Discussion and future directions}

We have constructed the spherically symmetric electron star solution in global AdS spacetime, by setting the fluid describing the star in full thermodynamic equilibrium with its environment, at finite temperature $T$ and chemical potential $\mu$. The first interesting result is that the previously described `electron cloud' solution~\cite{Hartnoll:2011,Thorlacius:2011} does not appear, as the thermodynamic equilibrium requires it to be fully swallowed by the black hole. Moreover, the spherical geometry provides an intrinsic length scale $a$ necessary to disentangle the temperature and chemical potential as independent thermodynamic dimensions. We interpret the resulting {\color{black}grand canonical phase diagram} as the dual description of a strongly coupled holographic metal at finite density confined within a spherical vessel.

We performed a Katz equilibrium analysis and found that the system possesses a {\color{black}thermodynamically} stable branch that eventually destabilizes at a critical central temperature, never regaining stability. This implies the existence of a finite region in the $(\tilde{\mu}_\infty, \tilde{T}_\infty)$ phase space where the star exists, extending up to a critical boundary where the {\color{black}two} 
branches meet. In contrast to the holographic neutron star case previously studied in~\cite{Grandi:2018}, the electron star exhibits a `pure core' structure exclusively when it is {\color{black}thermodynamically} stable, whereas the power-law asymptotic behavior emerges strictly once the solution becomes unstable.

We computed the grand canonical potential of the star and compared it with the Reissner-Nordström-AdS black hole (BH) and TAdS vacuum solutions within the saddle-point approximation. At finite temperature, the electron star is the dominant solution within a bounded thermodynamic region that depends strongly on the internal fluid parameters. For these regions, we mapped the corresponding phase transition curves separating the star from the black hole and TAdS phases. Depending on the choice of parameters, the electron star can undergo transitions to both the black hole and TAdS states, revealing a rich phase structure that includes triple points. Notably, these first-order phase transitions always occur well before the onset of the aforementioned second-order thermodynamic instability. Furthermore, we characterized the various quantum critical points at finite chemical potential, around which the global phase structure is organized.
 
Interestingly, for specific fluid parameters, the star-BH transition curve is qualitatively similar to the Hawking-Page curve obtained for the TAdS-BH vacuum phase transition~\cite{Chamblin:1999}. In the standard Hawking-Page scenario, this curve signals a confinement/deconfinement transition. 
In our case, we found another saddle point solution that dominates over TAdS and could serve as a novel background for constructing spherically symmetric holographic superconductors~\cite{Hartnoll:2008}, a possibility we plan to explore in future publications. 

A natural extension of this research would be to compute the bosonic and fermionic correlators on this background. As has already been done for the neutron star~\cite{Canavesi:2021,Acito:2024}, such an analysis is expected to exhibit rich holographic properties.

{\color{black}

Generally, a system whose entropy is a convex function of its variables is expected to be unstable against forming an inhomogeneous state. To see this, consider two adjacent identical subsystems with entropy \(S(N)\), where \(N\) is, for example, the number of particles. If the total entropy satisfies  
$
2S(N)< S(N+\delta N)+S(N-\delta N),
$ 
then moving particles from one subsystem to the other increases the entropy, so homogeneity is unstable. This instability corresponds to concavity of the grand canonical potential \(\Omega(\mu)\) as a function of the chemical potential \(\mu\).
The concave curves in Fig.~\ref{fig:FT0} might therefore suggest that this occurs in our system and that an inhomogeneous phase forms at the transition, as in an ice-water mixture at \(0^{\sf o}\)C. This would be consistent with the transition being first order: the derivative of \(\Omega\) with respect to \(\mu\), i.e. the particle number, is discontinuous at the transition. Thus, if one changes the particle number through the transition, a finite number of particles must be added or removed before one can move to the neighboring phase, analogous to latent heat in the ice-water case.

The above argument is not directly applicable here, however, because our system has finite volume. The expressions for the total entropy of the two subsystems may receive additional non-additive contributions from the interface between them. Such an interface contribution scales as \((\text{size})^2\), whereas the bulk contribution scales as \((\text{size})^3\), so it becomes negligible only in the infinite-volume limit. Therefore, {\color{black} the existence and properties of a mixed phase have not been demonstrated here and must be established separately}. A first step in such direction would be to calculate the phase diagram in the canonical ensemble, to compare it with the present grand canonical result (see \cite{referute1,referute2} for a similar program). Numerically, one could shoot from the asymptotic region inward and try to hit a regular solution at the center, or impose boundary conditions near the origin and near the boundary separately and then match the solutions at an intermediate radius.
}

\newpage
\section{Acknowledgements}

This work was partially supported by CONICET grant PIP-2023-11220220100262CO, and UNLP grant 2022-11/X931. The authors are grateful to Pablo Pisani, Adrián Lugo, Tobías Canavesi, Valentina Crespi, Carlos Arg\"uelles and Octavio Fierro for discussions on various subjects related to the present manuscript. L.A. thanks ICTP for hospitality and support during this work.
 
\appendix

\section*{A~~Details of the model}\label{app:star}
\addtocounter{section}{1}
We aim to describe the thermodynamics of a large number of charged self-gravitating fermions in equilibrium (also known as `Electron Stars') within a holographic framework. To this end, we consider a 3+1 dimensional global AdS spacetime and approximate the matter dynamics as a charged perfect fluid coupled to the gravitational field. The action functional consists of three contributions: the Einstein-Hilbert ($S_{\sf{Eins.}}$), Maxwell ($S_{\sf{Mxwl.}}$), and ideal fluid ($S_{\sf{Fluid}}$) terms
\begin{equation}\label{eq:S}
    S=\int d^4x \:\sqrt{-g}(\mathcal{L}_{ \sf Eins.}+\mathcal{L}_{\sf Mxwl.}+\mathcal{L}_{\sf Fluid})\,,
\end{equation}
where
\begin{eqnarray}
    \mathcal{L}_{\sf Eins.}&=&\frac{1}{2\kappa^2}\left(R+\frac{6}{L^2}\right)
    \\
    \mathcal{L}_{\sf Mxwl.}&=&-\frac{1}{4e^2}F_{\mu\nu}F^{\mu\nu}
    \\
    \mathcal{L}_{\sf Fluid}&=&-\rho(\sigma,s)+\sigma u^\mu(\partial_\mu\phi+\theta\:\partial_\mu s+A_\mu)+\lambda(u^\mu u_\mu+1)\,.   \label{eq:L_fluid}
\end{eqnarray}
We employ natural units ($\hbar=c=k_B=1$) and set the fermion charge to unity. The fluid sector follows the Schutz formalism \cite{Schutz:1970}. Here, $u^\mu$, $\rho$, $\sigma$, and $s$ denote the four-velocity, energy density, charge density, and entropy per particle of the fluid, respectively. Furthermore, $\lambda$ and $\theta$ are Lagrange multipliers, $\phi$ is a ``Clebsch" potential associated with the fluid velocity, and $A_\mu$ is the gauge potential (consequently, $\phi$ must shift under a gauge transformation to preserve gauge invariance). 

The equations of motion are obtained by taking functional variations of the action. Focusing on the fluid sector, variations with respect to $\delta\lambda$, $\delta\sigma$, $\delta s$, $\delta\theta$, and $\delta\phi$ yield
\begin{align}
    u^\mu u_\mu &= -1     
    \label{eq:timelike}
    \\
    \frac{\partial\rho(\sigma,s)}{\partial\sigma} &= u^\mu (\partial_\mu\phi+\theta\:\partial_\mu s+A_\mu)  
    \label{eq:mu1}
    \\
    \frac{\partial\rho(\sigma,s)}{\partial s} &= -\nabla_\mu(u^\mu\sigma\theta) 
    \label{eq:T1}
    \\
    \sigma u^{\mu}\partial_\mu s &=0 
    \label{eq:aux_s}
    \\
    \nabla_\mu(\sigma u^\mu) &= 0 \,. 
    \label{eq:continuity}
\end{align}
The first equation enforces the timelike normalization of the fluid velocity. If we identify the local chemical potential as
\begin{equation}
    \mu\equiv \frac{\partial\rho(\sigma,\theta)}{\partial \sigma}\,.
\end{equation}
Eq. \eqref{eq:mu1} can be rewritten as
\begin{equation}\label{eq:mu}
    \mu=u^\mu(\partial_\mu\phi+A_\mu)\,,
\end{equation}
where we used the entropy conservation condition \eqref{eq:aux_s} (characteristic of an adiabatic fluid) to eliminate the $u^\mu\partial_\mu s$ term. Notice that the chemical potential is gauge invariant and consists of the baryonic conservation term, $u^\mu\partial_\mu\phi$, plus the coupling to the gauge field, $u^\mu A_\mu$. For the Ansatz introduced in Sec.~\ref{sec:dynamics}, we set $\partial_\mu\phi=0$, such that the local chemical potential reads
\begin{equation}\label{eq:mu_star}
    \mu= \sqrt{-g^{00}}A_t= m\,\frac{h}{\tilde{m}\sqrt{f}}= m \,\tilde{\mu} \,,
\end{equation}
where $\tilde{\mu}=h/(\tilde{m}\sqrt{f})$ is the dimensionless chemical potential introduced in \eqref{eq:thermo_equilibrium}. Thus, the local chemical potential $\mu$ accounts for the local value of the background Maxwell field\footnote{One could also consider the case where $\partial_\mu\phi$ is a constant associated with particle number conservation. This is the case for neutron stars modeled as an uncharged perfect fluid \cite{Grandi:2018}. However, since the combination $\partial_\mu\phi+A_\mu$ is gauge invariant, we can set $\partial_\mu\phi=0$ without loss of generality, absorbing the constant into the gauge definition as the initial value $h(0)=\mu_0$.}.

Additionally, Eq. \eqref{eq:continuity} represents the continuity equation for the fluid current vector $J^\mu$ if we identify
\begin{equation}\label{eq:current}
    J^\mu\equiv\sigma u^\mu \,.
\end{equation}
This implies that \eqref{eq:aux_s} becomes the continuity equation for the total entropy, $\nabla_\mu(u^\mu\sigma s)=0$. Moreover, identifying the local temperature as
\begin{equation}
    \frac{\partial\rho}{\partial s}\equiv\sigma T \,,
\end{equation}
and utilizing the continuity equation \eqref{eq:continuity}, Eq. \eqref{eq:T1} simplifies to
\begin{equation}
    T=-u^\mu\partial_\mu\theta \,.
\end{equation}
Next, varying the action with respect to $\delta u^\mu$ leads to
\begin{equation}
    \sigma\,(\partial_\mu\phi+\theta\:\partial_\mu s+A_\mu)+2\,\lambda\, u_\mu=0\,.
\end{equation}
Contracting this equation with $u^\mu$ and using the previous results allows us to determine the Lagrange multiplier $\lambda$
\begin{equation}\label{eq:lambda}
    \lambda=\frac{\sigma u^\mu(\partial_\mu\phi+\theta\:\partial_\mu s+A_\mu)}{2}=\frac{\sigma \mu}{2}\,.
\end{equation}
On the other hand, varying the action with respect to the gauge potential $\delta A_\mu$ yields the Maxwell equations
\begin{equation}
    \nabla_\mu F^{\nu\mu}=e^2\,J^\nu\,.
\end{equation}
Finally, the variation with respect to the metric $\delta g_{\mu\nu}$ produces the Einstein equations with a negative cosmological constant $\Lambda=-3/L^2$, sourced by the energy-momentum tensors of the Maxwell field and the fluid
\begin{equation}\label{eq:Einstein_ap}
    R^{\mu\nu}-\frac{1}{2}g^{\mu\nu}\,R-\frac{3}{L^2}g^{\mu\nu}=\kappa^2\left(T_{\sf Mxwl.}^{\mu\nu} +T_{\sf Fluid}^{\mu\nu}\right)\,,
\end{equation}
where the energy-momentum tensors are derived from the Lagrangians as
\begin{equation}
    \begin{split}
        T^{\mu\nu}_{\sf Fluid} &=g^{\mu\nu}P+(\rho+P) u^{\mu} u^{\nu}
        \\
        T_{\sf Mxwl.}^{\mu\nu} &= \frac{1}{e^2}\left(F^{\mu\alpha}F^{\nu}_{\:\:\alpha}-\frac{1}{4}g^{\mu\nu}F_{\alpha\beta}F^{\alpha\beta}\right)\,.
    \end{split}
\end{equation}
We have thus derived the equations of motion for the model \eqref{eq:S}, which were introduced in Sec.~\ref{sec:dynamics} and solved in Sec.~\ref{sec:solutions} to obtain the Electron Star and vacuum solutions (Black Hole and TAdS).

As a result, we observe that the fluid sector of the action \eqref{eq:L_fluid} can be evaluated on-shell immediately by employing Eqs. \eqref{eq:timelike} and \eqref{eq:mu},
\begin{equation}\label{eq:fluid_onshell}
    \mathcal{L}_{\sf Fluid}^{\sf on-shell}=-\rho +\mu\,\sigma = \frac{\tilde{m}\,\tilde{\mu}\,\tilde{\sigma}-\tilde{\rho}}{L^2\kappa^2}\,,
\end{equation}
where in the last equality we applied the scaling notation introduced in Sec.~\ref{sec:dynamics}. Furthermore, we can use the local Gibbs-Duhem relation, $P+\rho = \mu\,\sigma + T\,S$, which remains locally valid in curved spacetime~\cite{Chavanis:2019}. In terms of the rescaled variables, this relation reads
\begin{equation}\label{eq:GibbsDuhem}
    \tilde{P}+\tilde{\rho} = \tilde{m}\,\tilde{\mu}\,\tilde{\sigma} + \tilde{T}\,\tilde{S}\,.
\end{equation}
Consequently, the on-shell fluid Lagrangian can be equivalently expressed as
\begin{equation}
    \mathcal{L}_{\sf Fluid}^{\sf on-shell}= P -T\,S= \frac{\tilde{P}-\tilde{T}\,\tilde{S}}{L^2\kappa^2}\,,
\end{equation}
with $S=s\,\sigma=\tilde{S}/m$ the local entropy density of the star. Notice that this is not the expected result for the variational principle of the Schutz model, as it differs by the $T\,S$ term. This term corresponds to a total derivative, $\nabla_\mu(\sigma \theta u^\mu s)=-T\,S$; thus, it does not affect the equations of motion but defines the boundary condition of the ensemble~\cite{Brown:1993}, which will be important when computing the grand canonical potential in Appendix B.

\newpage

\section*{B~~ Grand canonical potential computation details}\label{app:free_energy}
\addtocounter{section}{1}

In this appendix, we compute the on-shell action required for the grand canonical potential analysis presented in Sec.~\ref{sec:phase_transitions}. Since we compare saddle points sharing the same asymptotic symmetries, we first define the general procedure to compute the Euclidean on-shell action and subsequently apply it to each specific geometry. By performing a Wick rotation $t\rightarrow -i\,t_E$ on the Einstein action \eqref{eq:S}, with the Euclidean time periodicity $t_E \sim t_E + \beta$ (where $\beta=1/T$), we obtain
\begin{equation}\label{eq:S_def}
    I=-\int d^4x \:\sqrt{g}\,(\mathcal{L}_{ \sf Eins.}+\mathcal{L}_{\sf Mxwl.}+\mathcal{L}_{\sf Fluid})+I_{\sf GHY}+I_{\sf ct} \,,
\end{equation}
where $I_{\sf ct}$ is the holographic renormalization counterterm in asymptotically $AdS_4$ spaces \cite{Balasubramanian:1999},
\begin{equation}\label{eq:I_ct}
    I_{\sf ct}=\frac{1}{\kappa^2}\int d^3x \: \frac{2}{L}\sqrt{\gamma}\left(1+\frac{L^2}{4}R[\gamma]\right) \,.
\end{equation}
Here, $R[\gamma]$ is the Ricci scalar of the boundary metric $\gamma$ defined at the boundary hypersurface. On the other hand, $I_{\sf GHY}$ denotes the Gibbons-Hawking-York boundary term \cite{York:1972,Gibbons:1976},
\begin{equation}
    I_{\sf GHY}=-\frac{1}{\kappa^2} \int d^3x \:\sqrt{\gamma}\,K \,,
\end{equation}
written in terms of the extrinsic scalar curvature $K$.

The geometries of interest share the same asymptotic symmetries; thus, we can define the boundary terms generally and then compute them explicitly for each solution. We first define the unit normal vector $n^\mu$ to the hypersurface $\partial M$ at a constant cutoff $r=\Lambda$,
\begin{equation}
    n_\mu = \frac{\delta_\mu^r}{\sqrt{g^{rr}}} \,,
\end{equation}
where we have chosen the GH convention and defined the outgoing normal vector to the hypersurface. Then, the induced metric $\gamma_{\mu\nu}$ on the hypersurface $\partial M$ (the projection tensor) for a timelike unit normal vector $n^{\mu}$ is
\begin{equation}
    \gamma_{\mu\nu}=g_{\mu\nu}-n_\mu n_\nu \,.
\end{equation}
In our case, which involves static spherically symmetric metrics, the induced metric is simply $g_{ij}$ where $i,j$ run over the transverse directions. Finally, the extrinsic curvature $K_{\mu\nu}$ is given by the Lie derivative of the induced metric along the normal vector direction
\begin{equation}
    K_{\mu\nu}= \frac12\mathcal{L}_n\gamma_{\mu\nu}=\nabla_{(\mu}n_{\nu)} \qquad\text{and}\qquad K=g^{\mu\nu}K_{\mu\nu} \,.
\end{equation}

We are interested in comparing the free energies of different bulk geometries. To do so consistently, we must impose that they describe the same boundary spacetime at the same temperature. This is achieved by introducing a reference location at a cutoff $r=\Lambda$, where each system possesses a boundary thermal circle of the same proper length. The proper length of the thermal circle for a static observer along Euclidean time is defined as
\begin{equation}
    \ell_{\sf th}(r)= \in\tilde{T}_0^{T^{-1}}\sqrt{g_{00}(r)}\:dt_E = T^{-1}\sqrt{g_{00}(r)} \,.
\end{equation}
Thus, matching the thermal lengths at $r=\Lambda$, we require
\begin{equation}
    \tilde{\ell}_{\sf th}(\Lambda)=\ell_{\sf th}(\Lambda)\Longrightarrow \tilde{T}^{-1}\sqrt{\tilde{g}_{00}(\Lambda)}=T^{-1}\sqrt{g_{00}(\Lambda)} \,.
\end{equation}
Analogously, we match the chemical potentials. Since the chemical potential is defined as the gauge field component measured by a static observer, $u^\mu A_\mu=\mu$, we require the chemical potentials to coincide at the cutoff. In our framework, we have already defined the temperature and chemical potential such that, in the limit $\Lambda\rightarrow\infty$, we obtain
\begin{equation}\label{eq:T_mu_BHS}
    \mu_{\text{\tiny TAdS}}=\mu_{\text{\tiny BH}} = \tilde{\mu}_\infty\qquad\text{and}\qquad T_{\text{\tiny TAdS}}=T_{\text{\tiny BH}} = \tilde{T}_\infty \,.
\end{equation}
This ensures that the boundary conditions are identical for each solution. Consequently, we will express all quantities in terms of $\tilde{T}_\infty$ and $\tilde{\mu}_\infty$. It remains only to compute each grand canonical potential used in Sec.~\ref{sec:phase_transitions} to construct the {\color{black}grand canonical phase diagram}.

\paragraph{Black Hole.} We consider the Euclidean version of the Ansatz~\eqref{eq:Ansatz} for the black hole solution~\eqref{eq:bh_sol}, given by
\begin{equation}
    f(r)=g(r)^{-1}= \left(1-\frac{2M_0}{r}+\frac{Q_0^2}{2r^2}+r^2\right) \qquad\text{and}\qquad h(r)= -\frac{Q_0}{r} +\tilde{\mu}_\infty \,.
\end{equation}
The Hawking temperature is given by
\begin{equation}
    \tilde{T}_\infty=\frac{f'(r_0)}{4\pi} \,,
\end{equation}
where $r_0$ is the outer horizon radius. Recalling the gauge regularity condition $\tilde{\mu}_\infty=Q_0/r_0$, we can express the mass and charge in terms of $\tilde{T}_\infty$, $\tilde{\mu}_\infty$, and $r_0$
\begin{equation}
    Q_0=\tilde{\mu}_\infty\:r_0\qquad\text{and}\qquad M_0=\frac{r_0}2\left(\tilde{\mu}_\infty^2+4\pi\,r_0\, \tilde{T}_\infty-2\,r_0^2\right) \,.
\end{equation}
Moreover, substituting these into the definition of the outer horizon, it turns out that for fixed temperature and chemical potential, there are two possible horizon radii
\begin{equation}
    r_0^{\pm}= \frac{1}{6} \left(4\pi\,\tilde{T}_\infty\pm\sqrt{2}\sqrt{3 \,\tilde{\mu}_\infty^2 +8\pi^2\,\tilde{T}_\infty^2-6}\right) \,.
\end{equation}
These two branches, corresponding to the plus and minus signs, are referred to as the `large' and `small' black holes, respectively. Note that the existence of a physical horizon implies a constraint on $\tilde{T}_\infty$ and $\tilde{\mu}_\infty$,
\begin{equation}
    \begin{aligned}
        &T\geq\frac{\sqrt{\frac32(2-\tilde{\mu}_\infty^2)}}{2\pi} \qquad&&\text{and}\qquad |\tilde{\mu}_\infty|\leq \sqrt{2} \,,
        \\
        \text{or }\hspace{2cm}&&&
        \\
        \qquad &T\geq0\qquad&&\text{and}\qquad|\tilde{\mu}_\infty|>\sqrt{2}\qquad\text{(only for large BH)}\,.
    \end{aligned}
\end{equation}
The latter condition arises from requiring $M_0>0$. Since we work in the Grand Canonical Ensemble, both solutions contribute to the path integral; therefore, we sum over all allowed saddle points.

Let us compute the Black Hole Euclidean on-shell action using a radial cutoff $\Lambda$
\begin{equation}
   I_{\sf bulk} = -\in\tilde{T}_0^{\frac1{\tilde{T}_\infty}}dt_E\int d\Omega^2\int_{r_0^\pm}^{\Lambda}dr\:\sqrt{g}\left(\frac{1}{2\kappa^2}\left(R+\frac{6}{L^2}\right)-\frac{1}{4e^2}F_{\mu\nu}F^{\mu\nu}\right) \,.
\end{equation}
Using the on-shell relations $R=-12/L^2$ and $F^2=-2e^2(\tilde{\mu}_\infty\:r_0^{\pm})^2/(\kappa^2L^2r^4)$, we obtain
\begin{equation}
    I_{\sf bulk} = \frac{2\pi\,L^2}{\kappa^2\,\tilde{T}_\infty} \left(2 \Lambda ^3 -r_0^\pm\left(2(r_0^\pm)^2  +\tilde{\mu}_\infty^2\right) +\tilde{\mu}_\infty^2\frac{(r_0^\pm)^2}{\Lambda}\right)\,.
\end{equation}
Notice that this term exhibits a cubic divergence $\mathcal{O}(\Lambda^{3})$ as $\Lambda\rightarrow\infty$.
\\
The GHY term evaluates to
\begin{equation}
    I_{\sf GHY}= \frac{2\pi\,L^2}{\kappa^2\,\tilde{T}_\infty} \left(3\,r_0^\pm \left(\tilde{\mu}_\infty^2 -2(r_0^\pm)^2 +4\pi\,r_0^\pm\,\tilde{T}_\infty\right) -\tilde{\mu}_\infty^2\frac{(r_0^\pm)^2}{\Lambda} -4\Lambda -6\Lambda^3\right) \,.
\end{equation}
This term contains divergences of order $\mathcal{O}(\Lambda^{3})$ and $\mathcal{O}(\Lambda^{1})$.
\\
Finally, we compute the counterterm \eqref{eq:I_ct} to regularize the total action 
\begin{equation}
    \begin{aligned}
        I_{\sf ct}&=\frac{2\pi\,L^2}{\kappa^2\,\tilde{T}_\infty}\frac{\left(2 \Lambda ^2+1\right)}{\Lambda} \sqrt{4\Lambda\left(\Lambda^3 +\Lambda +2(r_0^\pm)^3-\tilde{\mu}_\infty^2 \,r_0^\pm\right) +2(r_0^\pm)^2 \left(\tilde{\mu}_\infty^2 -8\pi\,\Lambda\,\tilde{T}_\infty\right)}
        \\
        &\approx \frac{4\pi\,L^2}{\kappa^2\,\tilde{T}_\infty} \left(2\Lambda^3 +2\Lambda +r_0^\pm\left(2(r_0^\pm)^2 -4\pi\,r_0^\pm\,\tilde{T}_\infty -\tilde{\mu}_\infty^2\right)\right) +\mathcal{O}(\Lambda^{-1}) \,.
    \end{aligned}
\end{equation}
All divergences cancel out exactly when combining these terms. Thus, taking the limit $\Lambda\rightarrow\infty$, the total regularized action is
\begin{equation}
    I_{\text{\tiny BH}}^{\pm}= \frac{8\pi\,L^2}{\kappa^2\,\tilde{T}_\infty} (r_0^\pm)^2 (\pi\,\tilde{T}_\infty -r_0^\pm) \,.
\end{equation}
Since $r_0^+\geq r_0^-$, it follows that 
\begin{equation}
    I^{+}_{\text{\tiny BH}}\leq I^{-}_{\text{\tiny BH}}
\end{equation}
Consequently, the grand canonical potential of the large black hole is always lower than that of the small black hole, making it the dominant contribution. For the construction of the {\color{black}grand canonical phase diagram}, we will consider only the large black hole branch. Its grand canonical potential is given by \eqref{eq:free_energies}.

\paragraph{Thermal AdS$_4$.} The TAdS grand canonical potential is obtained immediately by taking the limit $M_0=Q_0=0$ (implying $r_0=0$) in the black hole calculation. This yields $F_{\sf TAdS}=0$ for all $\tilde{T}_\infty$ and $\tilde{\mu}_\infty$. Comparing the TAdS potential with that of the charged black hole leads to the Hawking-Page phase transition curve \cite{Chamblin:1999}. 

\paragraph{Electron Star.} The details of the Electron Star model are provided in Appendix A. The Euclidean action is given by
\begin{equation}
    I_{\sf bulk}=-\int d^4x \:\sqrt{g}\left(\frac{1}{2\kappa^2}\left(R+\frac{6}{L^2}\right)-\frac{1}{4e^2}F_{\mu\nu}F^{\mu\nu} +\frac{\tilde{m}\,\tilde{\mu}\,\tilde{\sigma}-\tilde{\rho}}{\kappa^2L^2}\right) \,,
\end{equation}
where the fluid term has already been evaluated on-shell using \eqref{eq:fluid_onshell}. Upon substituting the equations of motion \eqref{eq:eom_star}, the integrand reduces entirely to a total derivative. Since the star solution is required to be regular at the origin (as detailed in Sec.~\ref{sec:dynamics}), the action reduces to a boundary term at the cutoff $\Lambda$,
\begin{equation}
    I_{\sf bulk} = \frac{2\pi \,L^2}{\kappa ^2
   \tilde{T}_\infty}\,\Lambda^2\, e^{-\frac{\chi (\Lambda)}{2}} \left(f'(\Lambda) -2\,h'(\Lambda)\,(\tilde{m}\,\tilde{\mu}_0+h(\Lambda)) \right) \,.
\end{equation}
Next, we compute the GHY term
\begin{equation}
    I_{GHY}= 
    -\frac{2\pi\,L^2}{\kappa^2\,\tilde{T}_\infty}e^{-\frac{\chi(\Lambda)}{2}}\,\Lambda \left(\Lambda\,f'(\Lambda) +4\,f(\Lambda )\right) \,.
\end{equation}
Finally, the counterterm contribution \eqref{eq:I_ct} is
\begin{equation}
    I_{\sf ct} = \frac{4\pi\,L^2}{\kappa ^2 \tilde{T}_\infty} \left(2 \Lambda ^2+1\right)\,f(\Lambda)^{\frac12} \,,
\end{equation}
Combining these contributions, taking the cutoff $\Lambda\rightarrow\infty$, and using the asymptotic form of the metric \eqref{star_asymp}, the regularized Euclidean action yields
\begin{equation}\label{eq:ES_action}
    I^{(\sf Star)}=
    \frac{4\pi\,L^2}{\kappa^2 \tilde{T}_\infty}\,e^{\frac{\chi_s}{2}}\left(2 \,M_s-\tilde{\mu}_\infty\,Q_s\right) \,.
\end{equation}
From the discussion in the Appendix A, this action yields the thermodynamic potential associated with the adiabatic constraints of the fluid. To recover the Grand Canonical Potential $\Omega$, we must correct the boundary term, which is just a Legendre transformation relating to entropy:
\begin{equation}
    \Omega= \tilde{T}_\infty\,I^{(\sf Star)} -\tilde{T}_{\infty}\,\tilde{S}_{s} \,,
\end{equation}
where $\tilde{S}_s$ is the total entropy of the star~\cite{Thorlacius:2011}. This correction is only need for the Electron Star since the entropic origin is different: the on-shell black hole action gives correctly the grand canonical energy since it entropy comes from the horizon; while in the star the geometric part has zero entropy, thus it can only comes from the star which we first computed for a adiabatic fluid and then transform back. The entropy can be computed integrating spatially the entropy density \cite{Chavanis:2008}, which using the Gibbs-Duhem relation \eqref{eq:GibbsDuhem} combined with the local temperature and chemical potential satisfying the equilibrium condition \eqref{eq:thermo_equilibrium}, yields
\begin{equation} 
    \tilde{S}_s=\frac{4\pi \,L^2}{\kappa^2}\int_0^\infty dr\,r^2\sqrt{g(r)}\,\tilde{S}= \frac{4\pi\,L^2}{\kappa^2\,\tilde{T}_0}\int_0^\infty dr\,\,r^2 e^{\frac{\chi}{2}}\left(\tilde{\rho}+\tilde{P}-\frac{h}{\sqrt{f}}\,\tilde{\sigma}\right) \,.   
\end{equation}
This concludes the derivation of the free energies for the three classical geometries used in \eqref{eq:free_energies} to construct the complete {\color{black}grand canonical phase diagram} in Sec.~\ref{sec:phase_transitions}.

\newpage

\section*{C~~ Black hole surrounded by an electron cloud}\label{app:electron_cloud}
\addtocounter{section}{1}

In this Appendix, we analyze a Reissner-Nordström black hole surrounded by a gas of charged particles, referred to as an \textit{electron cloud}. This type of configuration has been studied at zero temperature in~\cite{Hartnoll:2011}, and with temperature effects in~\cite{Thorlacius:2011}. However, in the latter cases, the cloud is not in thermodynamic equilibrium with the black hole but is instead fixed at a background temperature of $T=0$. We argue here that introducing a finite temperature to construct an electron cloud in full thermodynamic equilibrium with the black hole is not possible within this framework.

We obtained the black hole solution in Sec.~\ref{sec:solutions}. From the regularity conditions, the black hole temperature and gauge potential are given by
\begin{equation}
    T_{\text{\tiny BH}}=\frac{f'(r_0)}{4\pi} \qquad\text{and}\qquad h(r)=-\frac{Q_0}{r}+\mu_{\text{\tiny BH}} \,,
\end{equation}
where $\mu_{\text{\tiny BH}}=Q_0/r_0$, $r_0$ is the outer horizon radius, and $f(r)$ and $h(r)$ are the black hole metric functions~\eqref{eq:bh_sol}. On the other hand, the particle cloud is treated as a perfect fluid following the model introduced in Appendix A. Due to the spherical symmetry of the black hole, the cloud must respect this symmetry. A necessary and sufficient condition for the fluid to be in global thermal equilibrium is that it satisfies the local Tolman and gauge conditions 
\begin{equation}
    T(r)=\frac{\tilde{T}_0}{\sqrt{f(r)}} \qquad\text{and}\qquad \mu(r)=\frac{h(r)}{\sqrt{f(r)}} \,,
\end{equation}
where $\tilde{T}_0$ and $h(r_{\sf min})=\tilde{\mu}_0$ are the initial conditions for the cloud at the inner border of the cloud $r_{\sf min}$. We use these conditions to solve the equations of state numerically. For simplicity, we focus the analysis on the Fermi-Dirac integrals of the form
\begin{equation}
    \mathcal{I}=\int_1^{\infty} \frac{H(\epsilon)}{e^{\frac{(\epsilon-\mu)}{T}}+1}\:d\epsilon \,,
\end{equation}
where $H(\epsilon)$ represents the ``density of states'' function appearing in the definitions of $\rho$, $\sigma$, and $P$ in~\eqref{eq:thermodynamics}. 

Notice that attempting to construct a thermodynamically stable solution for a cloud extending down to the black hole horizon leads to a divergence. If we impose thermal equilibrium with the black hole by setting $\tilde{T}_0=T_{\text{\tiny BH}}$ (\textit{i.e.} $r_{\sf min}=r_0$) the local temperature $T(r)$ diverges as $r \to r_0$ (since $f(r_0)=0$). Consequently, the local fluid density and pressure would diverge at the horizon. The only consistent solution in this setup is to set $\tilde{T}_0=0$, \textit{i.e.}, the electron cloud must be at zero temperature (a fully degenerate Fermi gas). In this scenario, the fluid density vanishes at the horizon and the cloud only exists in regions where the local chemical potential satisfies $\mu(r) \geq 1$. This results in a black hole surrounded by a zero-temperature cloud, potentially with a global temperature parameter shifted by gravitational backreaction, but not in thermal contact with the horizon~\cite{Thorlacius:2011}. 

\begin{figure}
	\begin{center}
	\includegraphics[width=0.49\textwidth]{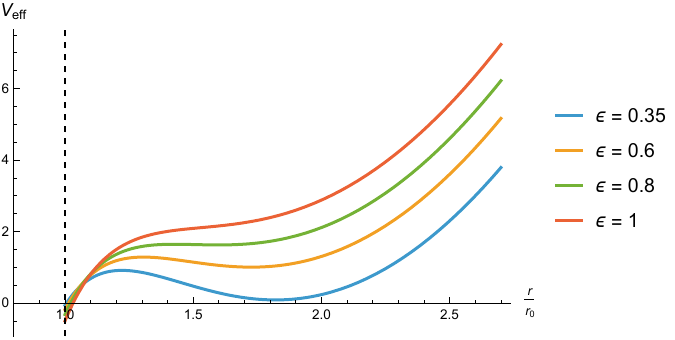}%
    \hfill \includegraphics[width=0.49\textwidth]{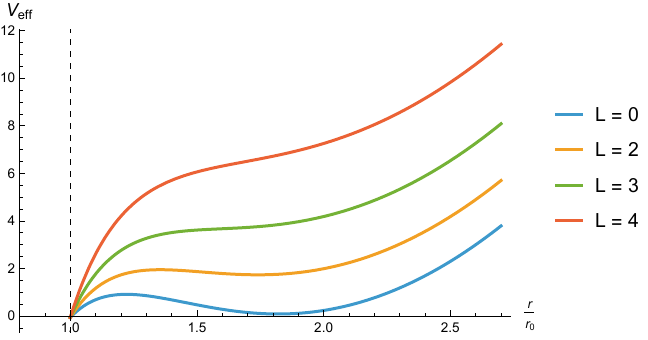}
		\caption{Plots of the effective potential \eqref{eq:Veff}, fixing the angular momentum $L=0$ (Left) and the energy $\epsilon=0.35$ (Right). Both plots has the horizon fixed (black dash line) at $M=10$ and $Q=4.5$.
        Plots of the effective potential \eqref{eq:Veff}. \textbf{Left:} Fixed angular momentum $L=0$ with varying energy. \textbf{Right:} Fixed energy $\epsilon=0.35$ with varying angular momentum. Both plots assume a fixed background with horizon radius corresponding to $M=10$ and $Q=4.5$, and the charge and mass of the particle are $q=1$ and $m=0.2$ respectively.}
        \label{fig:geodesics}
	\end{center}
\end{figure}

While static thermal equilibrium is forbidden, dynamically stable geodesics exist. The timelike condition $g_{\mu\nu}\dot{x}^\mu\dot{x}^\nu=-1$ together with the constants of motion implies $\frac{1}{2}\dot{r}^2 - V_{\text{eff}}(r) = 0$, with the effective potential~\cite{Blau:2025}
\begin{equation}\label{eq:Veff}
    V_{\text{eff}}(r)=\frac{f(r)}{2}\left(1+\frac{L^2}{2}\right)-\frac12\left(\epsilon+\frac{q}{m}\,h(r)\right)^2 \,,
\end{equation}
with $q$, $m$ and $L$ being the charge, mass and angular momentum of the particle. Note that this potential depends on the energy $\epsilon$ and cannot be decoupled; allowed trajectories must satisfy $V_{\text{eff}}(r) \ge 0$. As shown in Fig.~\ref{fig:geodesics}, the potential possesses stable minima. If a particle remains at the corresponding energy, the dynamics are stable; statistically, this corresponds to a zero-temperature cloud. However, introducing a finite temperature implies thermal fluctuations. These fluctuations allow particles in stable orbits to gain energy, overcome the potential barrier, and eventually fall into the horizon. This process continues until the cloud is depleted or reaches a fully degenerate ($T=0$) state.

In conclusion, an electron cloud can only be considered at zero temperature and cannot coexist in thermal equilibrium with a finite-temperature black hole in this formalism. This is consistent with the fact that the equations of state used for the electron star assume a vacuum state where the absence of particles corresponds to zero temperature (Boulware vacuum); consequently, the horizon is a fixed boundary with no radiation. The divergence at the horizon could be regularized using a Hartle-Hawking vacuum, which defines thermodynamic equilibrium including Hawking radiation~\cite{Candelas:1980}. In that case, the horizon is dynamic and could exchange heat with the cloud. However, since we are interested in holographic applications, we restrict our analysis to the Boulware vacuum, as this corresponds to the ground state in the dual CFT, whereas the Hartle-Hawking vacuum corresponds to a thermal state in the dual theory~\cite{Witten:1998}.


\newpage

\begin{thebibliography}{99}
    
    \bibitem{Hartnoll:2008} S.~A.~Hartnoll, C.~P.~Herzog and G.~T.~Horowitz, \emph{Building a Holographic Superconductor}, \emph{Phys. Rev. Lett.} \textbf{101} (2008) 031601 [\href{https://arxiv.org/abs/0803.3295}{{\tt arXiv:0803.3295}}].
    
    \bibitem{Faulkner:2010} T.~Faulkner, G.~T.~Horowitz, J.~McGreevy, M.~M.~Roberts and D.~Vegh, \emph{Photoemission 'experiments' on holographic superconductors}, \emph{JHEP} \textbf{03} (2010) 121 [\href{https://arxiv.org/abs/0911.3402}{{\tt arXiv:0911.3402}}].

    \bibitem{Gubser:2010} S.~S.~Gubser, F.~D.~Rocha and P.~Talavera, \emph{Normalizable fermion modes in a holographic superconductor}, \emph{JHEP} \textbf{10} (2010) 087 [\href{https://arxiv.org/abs/0911.3632}{{\tt arXiv:0911.3632}}].
    
    \bibitem{Hartnoll:2004} S.~A.~Hartnoll, J.~Polchinski, E.~Silverstein and D.~Tong, \emph{Towards strange metallic holography}, \emph{JHEP} \textbf{04} (2010) 120 [\href{https://arxiv.org/abs/0912.1061}{{\tt arXiv:0912.1061}}].

    \bibitem{Faulkner:2011} T.~Faulkner, H.~Liu, J.~McGreevy and D.~Vegh, \emph{Emergent quantum criticality, Fermi surfaces, and AdS$_2$}, \emph{Phys. Rev. D} \textbf{83} (2011) 125002 [\href{https://arxiv.org/abs/0907.2694}{{\tt arXiv:0907.2694}}].

    \bibitem{Cubrovic:2009} M.~Cubrovic, J.~Zaanen and K.~Schalm, \emph{String Theory, Quantum Phase Transitions and the Emergent Fermi-Liquid}, \emph{Science} \textbf{325} (2009) 439 [\href{https://arxiv.org/abs/0904.1993}{{\tt arXiv:0904.1993}}].

    \bibitem{Hartnoll:2011} S.~A.~Hartnoll and A.~Tavanfar, \emph{Electron stars for holographic metallic criticality}, \emph{Phys. Rev. D} \textbf{83} (2011) 046003 [\href{https://arxiv.org/abs/1008.2828}{{\tt arXiv:1008.2828}}].

    \bibitem{Hartnoll:2011b} S.~A.~Hartnoll, D.~M.~Hofman and A.~Tavanfar, \emph{Holographically smeared Fermi surface: Quantum oscillations and Luttinger count in electron stars}, \emph{Europhys. Lett.} \textbf{95} (2011) 31002 [\href{https://arxiv.org/abs/1011.2502}{{\tt arXiv:1011.2502}}].

    \bibitem{Kiritsis:2015hoa}
    E.~Kiritsis and L.~Li,
    \emph{``Holographic Competition of Phases and Superconductivity,''}
    JHEP \textbf{01} (2016), 147
    doi:10.1007/JHEP01(2016)147
    [arXiv:1510.00020 [cond-mat.str-el]].

    \bibitem{Verlinde:2011} J.~de Boer, K.~Papadodimas and E.~Verlinde, \emph{Holographic Neutron Stars}, \emph{JHEP} \textbf{10} (2010) 020 [\href{https://arxiv.org/abs/0907.2695v3}{{\tt arXiv:0907.2695}}];
    
    \bibitem{Arsiwalla:2011}
    X.~Arsiwalla, J.~de Boer, K.~Papadodimas and E.~Verlinde, \emph{Degenerate Stars and Gravitational Collapse in AdS/CFT}, \emph{JHEP} \textbf{01} (2011) 144 [\href{https://arxiv.org/abs/1010.5784}{{\tt arXiv:1010.5784}}]

    \bibitem{Grandi:2018} C.~A.~Arg\"uelles and N.~Grandi, \emph{Fermionic halos at finite temperature in AdS/CFT}, \emph{JHEP} \textbf{05} (2018) 118 [\href{https://arxiv.org/abs/1712.05866}{{\tt arXiv:1712.05866}}].

    \bibitem{Grandi:2020} C.~A.~Arg\"uelles, E.~Canavesi, M.~Diaz and N.~Grandi, \emph{Thermodynamic instabilities in holographic neutron stars at finite temperature}, \emph{Class. Quant. Grav.} \textbf{37} (2020) 205002 [\href{https://arxiv.org/abs/1911.02554}{{\tt arXiv:1911.02554}}].

    \bibitem{Canavesi:2023} E.~Canavesi, \emph{Holographic neutron stars at finite temperature}, [\href{https://arxiv.org/abs/2312.10021}{{\tt arXiv:2312.10021}}].

    \bibitem{Arguelles:2013} Arg\"uelles, C., Siutsou, I., Ruffini, R., Rueda, J.,  Machado, B. \emph{On the core-halo constituents of a semi-degenerate gas of massive fermions}. Probes of Dark Matter on Galaxy Scales (2013), 1, 30204.


    \bibitem{Arguelles:2014}  Arg\"ulles, C. R., Ruffini, R., Siutsou, I., Fraga, B. \emph{On the distribution of dark matter in galaxies: Quantum treatments.} Journal of the Korean Physical Society (2014), 65(6), 801-804.

    \bibitem{Ruffini:2015} Ruffini, R., Argüelles, C. R.,  Rueda, J. A. \emph{On the core-halo distribution of dark matter in galaxies}. Monthly Notices of the Royal Astronomical Society (2015), 451(1), 622-628.

    \bibitem{Hartnoll:2010} S.~A.~Hartnoll and P.~Petrov, \emph{Electron Star Birth: A Continuous Phase Transition at Nonzero Density}, \emph{Phys. Rev. Lett.} \textbf{106} (2011) 121601 [\href{https://arxiv.org/abs/1011.6469}{{\tt arXiv:1011.6469}}].

    \bibitem{Thorlacius:2011} V.~G.~M.~Puletti, S.~Nowling, L.~Thorlacius and T.~Zingg, \emph{Holographic metals at finite temperature}, \emph{JHEP} \textbf{01} (2011) 117 [\href{https://arxiv.org/abs/1011.6261}{{\tt arXiv:1011.6261}}].



    \bibitem{Balasubramanian:1999} V.~Balasubramanian and P.~Kraus, \emph{A stress tensor for Anti-de Sitter gravity}, \emph{Commun. Math. Phys.} \textbf{208} (1999) 413 [\href{https://arxiv.org/abs/hep-th/9902121}{{\tt hep-th/9902121}}].

    \bibitem{Gibbons:1976} G.~W.~Gibbons and S.~W.~Hawking, \emph{Action Integrals and Partition Functions in Quantum Gravity}, \emph{Phys. Rev. D} \textbf{15} (1977) 2752.

    \bibitem{York:1972} J.~W.~York, Jr., \emph{Role of conformal three-geometry in the dynamics of gravitation}, \emph{Phys. Rev. Lett.} \textbf{28} (1972) 1082.

    \bibitem{Katz:1978} J.~Katz, \emph{On the number of unstable modes of an equilibrium}, \emph{Mon. Not. Roy. Astron. Soc.} \textbf{183} (1978) 765.

    \bibitem{Katz:2002} J.~Katz, \emph{Thermodynamics and selfgravitating systems}, \emph{Found. Phys.} \textbf{33} (2003) 223 [\href{https://arxiv.org/abs/astro-ph/0212295}{{\tt astro-ph/0212295}}].

    \bibitem{Chavanis:2019} P.-H.~Chavanis, \emph{Statistical mechanics of self-gravitating systems in general relativity: I. The quantum Fermi gas}, \emph{Eur. Phys. J. Plus} \textbf{135} (2020) 290 [\href{https://arxiv.org/abs/1908.10806}{{\tt arXiv:1908.10806}}].

    \bibitem{Kobayashi:2007} S.~Kobayashi, D.~Mateos, S.~Matsuura, R.~C.~Myers and R.~M.~Thomson, \emph{Holographic phase transitions at finite baryon density}, \emph{JHEP} \textbf{02} (2007) 016 [\href{https://arxiv.org/abs/hep-th/0611099}{{\tt hep-th/0611099}}].

    \bibitem{Chamblin:1999} A.~Chamblin, R.~Emparan, C.~V.~Johnson and R.~C.~Myers, \emph{Charged AdS black holes and catastrophic holography}, \emph{Phys. Rev. D} \textbf{60} (1999) 064018 [\href{https://arxiv.org/abs/hep-th/9902170}{{\tt hep-th/9902170}}].

    \bibitem{Kapusta:2006} J.~I.~Kapusta and C.~Gale, \emph{Finite-temperature field theory: Principles and applications}, Cambridge University Press (2006).

    \bibitem{Glendenning:1997} N.~K.~Glendenning, \emph{Compact stars: Nuclear physics, particle physics, and general relativity}, Springer (1997).

    \bibitem{Bekenstein:1971} J.~D.~Bekenstein, \emph{Hydrostatic Equilibrium and Gravitational Collapse of Relativistic Charged Fluid Balls}, \emph{Phys. Rev. D} \textbf{4} (1971) 2185.

    \bibitem{Landau:1980} L.~D.~Landau and E.~M.~Lifshitz, \emph{Statistical Physics, Part 1}, Butterworth-Heinemann (1980).

    \bibitem{Shi:2021} K.~Shi, Y.~Tian, X.~Wu, H.~Zhang and C.~Zhu, \emph{Thermodynamic equilibrium condition and the first law of thermodynamics for charged perfect fluids in electromagnetic and gravitational fields}, \emph{Class. Quant. Grav.} \textbf{39} (2022) 085004 [\href{https://arxiv.org/abs/2108.08729v2}{{\tt arXiv:2108.08729}}].

    \bibitem{Schutz:1970} B.~F.~Schutz, \emph{Perfect Fluids in General Relativity: Velocity Potentials and a Variational Principle}, \emph{Phys. Rev. D} \textbf{2} (1970) 2762.

    \bibitem{Github} The code used to compute the electron star profiles, perform the Katz analysis, and calculate the grand canonical potenital can be found in this GitHub repository: \href{https://github.com/acitolucas/holographic-electron-stars.git}{\tt https://github.com/acitolucas/holographic-electron-stars.git}.

    \bibitem{Brown:1993} J.~D.~Brown, \emph{Action functionals for relativistic perfect fluids}, \emph{Class. Quant. Grav.} \textbf{10} (1993) 1579 [\href{https://arxiv.org/abs/gr-qc/9304026}{{\tt gr-qc/9304026}}].
    
    \bibitem{Chavanis:2008} P.~H.~Chavanis, \emph{Relativistic stars with a linear equation of state: analogy with classical isothermal spheres and black holes}, \emph{Astron. Astrophys.} \textbf{483} (2008) 673 [\href{https://arxiv.org/abs/0707.2292}{{\tt arXiv:0707.2292}}].

    \bibitem{Canavesi:2021} E.~Canavesi, O.~Fierro, N.~Grandi and P.~Pisani, \emph{Scalar correlators and normal modes in holographic neutron stars}, \emph{Class. Quant. Grav.} \textbf{40} (2023) 025001 [\href{https://arxiv.org/abs/2205.04374}{{\tt arXiv:2205.04374}}].

    \bibitem{Acito:2024} M.~Acito, E.~Canavesi, N.~Grandi and A.~Lugo, \emph{Fermionic correlators on the holographic neutron star}, \emph{JHEP} \textbf{04} (2024) 153 [\href{https://arxiv.org/abs/2401.03362}{{\tt arXiv:2401.03362}}].

    \bibitem{Candelas:1980} P.~Candelas, \emph{Vacuum Polarization in Schwarzschild Spacetime}, \emph{Phys. Rev. D} \textbf{21} (1980) 2185.

    \bibitem{Witten:1998} E.~Witten, \emph{Anti-de Sitter space and holography}, \emph{Adv. Theor. Math. Phys.} \textbf{2} (1998) 253 [\href{https://arxiv.org/abs/hep-th/9802150}{{\tt hep-th/9802150}}].

    \bibitem{Blau:2025} M.~Blau, \emph{Lecture Notes on General Relativity}, University of Bern (2025), \href{http://www.blau.itp.unibe.ch/GRLecturenotes.html}{http://www.blau.itp.unibe.ch/GRLecturenotes.html}. 

    \bibitem{referute1} D.~Mateos, S.~Matsuura, R.~Myers and R.~Thomson, \emph{Holographic phase transitions at finite chemical potential}, \emph{JHEP} \textbf{11} (2007) 085 [\href{https://iopscience.iop.org/article/10.1088/1126-6708/2007/11/085}{{\tt 10.1088/1126-6708/2007/11/085}}]

    \bibitem{referute2} K.~Shinpei, D.~Mateos, S.~Matsuura, R.~Myers and R.~Thomson, \emph{Holographic phase transitions at finite baryon density}, \emph{JHEP} \textbf{02} (2007) 016 [\href{https://iopscience.iop.org/article/10.1088/1126-6708/2007/02/016/meta?paper=02%282007%29016}{{\tt 10.1088/1126-6708/2007/02/016}}]
    
\end{thebibliography}
\end{document}